\documentclass[11pt, a4paper]{article}

\usepackage{jheppub, slashed, color,subfig}
\usepackage{enumitem}
\usepackage{multirow}
\usepackage{tikz}
\usetikzlibrary{decorations.pathmorphing}
\usepackage{tikz-3dplot}

\usepackage{amsmath}
\allowdisplaybreaks[2]
\usepackage{color}
\usepackage[multiple]{footmisc}

\usepackage{textcomp}
\usepackage{amssymb}
\usetikzlibrary{positioning}
\usetikzlibrary{calc,intersections,through,backgrounds}

\usepackage{tikz-3dplot}
\usepackage{amscd}
\usepackage{amsthm}
\usepackage{array} 
\usepackage{graphicx}
\usepackage{adjustbox}
\usepackage{mathrsfs}
\usepackage{dsfont}
\newcommand{\C}{\mathbb{C}}

\DeclareFontFamily{U}{MnSymbolC}{}
\DeclareSymbolFont{MnSyC}{U}{MnSymbolC}{m}{n}
\DeclareFontShape{U}{MnSymbolC}{m}{n}{
    <-6>  MnSymbolC5
   <6-7>  MnSymbolC6
   <7-8>  MnSymbolC7
   <8-9>  MnSymbolC8
   <9-10> MnSymbolC9
  <10-12> MnSymbolC10
  <12->   MnSymbolC12}{}
\DeclareMathSymbol{\intprod}{\mathbin}{MnSyC}{'270}

\newcommand{\R}{\mathbb{R}}

\newcommand{\til}{\widetilde}

\newcommand{\zb}{\bar{z}}

\newcommand{\bR}{\mathbb{R}}

\newcommand{\D}{\delta}

\let\nc\newcommand
\let\renc\renewcommand
\nc{\wbar}{\overline}
\let\td\tilde
\let\wtd\widetilde
\let\wht\widehat
\let\mcl\mathcal

\nc{\ab}{{\bar{a}}} \nc{\at}{\tilde{a}} \nc{\ah}{\hat{a}}
\nc{\bb}{{\bar{b}}} 
\nc{\cb}{{\bar{c}}} \nc{\ct}{\tilde{c}} 
\nc{\fb}{{\bar{f}}} \nc{\ft}{\tilde{f}} \nc{\fh}{\hat{f}}
\nc{\gb}{{\bar{g}}} \nc{\gt}{\tilde{g}} \nc{\gh}{\hat{g}}
\nc{\hb}{{\bar{h}}} \nc{\hh}{\hat{h}} 
\nc{\ib}{{\bar{\imath}}} \nc{\ih}{\hat{\imath}} 
\nc{\jb}{{\bar{\jmath}}} \nc{\jt}{\tilde{\jmath}} \nc{\jh}{\hat{\jmath}}
\nc{\kb}{{\bar{k}}} \nc{\kt}{\tilde{k}} \nc{\kh}{\hat{k}}
\nc{\lb}{{\bar{l}}} \nc{\lt}{\tilde{l}} \nc{\lh}{\hat{l}}
\nc{\mb}{{\bar{m}}} \nc{\mt}{\tilde{m}} \nc{\mh}{\hat{m}}
\nc{\nb}{{\bar{n}}} \nc{\nt}{\tilde{n}} \nc{\nh}{\hat{n}}
\nc{\ob}{{\bar{o}}} \nc{\ot}{\tilde{o}} \nc{\oh}{\hat{o}}
\nc{\pb}{{\bar{p}}} \nc{\pt}{\tilde{p}} \nc{\ph}{\hat{p}}
\nc{\qb}{{\bar{q}}} \nc{\qt}{\tilde{q}} \nc{\qh}{\hat{q}}
\nc{\rb}{{\bar{r}}} \nc{\rt}{\tilde{r}} 
\renc{\sb}{{\bar{s}}} \nc{\st}{\tilde{s}} \nc{\sh}{\hat{s}}
\nc{\tb}{{\bar{t}}} \renc{\th}{\hat{t}} 
\nc{\ub}{{\bar{u}}} \nc{\ut}{\tilde{u}} \nc{\uh}{\hat{u}}
\nc{\vb}{{\bar{v}}} \nc{\vt}{\tilde{v}} \nc{\vh}{\hat{v}}
\nc{\wb}{{\bar{w}}} \nc{\wt}{\tilde{w}} \nc{\wh}{\hat{w}}
\nc{\xb}{{\bar{x}}} \nc{\xt}{\tilde{x}} \nc{\xh}{\hat{x}}
\nc{\yb}{{\bar{y}}} \nc{\yt}{\tilde{y}} \nc{\yh}{\hat{y}}

\nc{\Ab}{\wbar{A}} \nc{\At}{\wtd{A}} \nc{\Ah}{\wht{A}}
\nc{\Bb}{\wbar{B}} \nc{\Bt}{\wtd{B}} \nc{\Bh}{\wht{B}}
\nc{\Cb}{\wbar{C}} \nc{\Ct}{\wtd{C}} \nc{\Ch}{\wht{C}}
\nc{\Db}{\wbar{D}} \nc{\Dt}{\wtd{D}} \nc{\Dh}{\wht{D}}
\nc{\Eb}{\wbar{E}} \nc{\Et}{\wtd{E}} \nc{\Eh}{\wht{E}}
\nc{\Fb}{\wbar{F}} \nc{\Ft}{\wtd{F}} \nc{\Fh}{\wht{F}}
\nc{\Gb}{\wbar{G}} \nc{\Gt}{\wtd{G}} \nc{\Gh}{\wht{G}}
\nc{\Hb}{\wbar{H}} \nc{\Ht}{\wtd{H}} \nc{\Hh}{\wht{H}}
\nc{\Ib}{\wbar{I}} \nc{\It}{\wtd{I}} \nc{\Ih}{\wht{I}}
\nc{\Jb}{\wbar{J}} \nc{\Jt}{\wtd{J}} \nc{\Jh}{\wht{J}}
\nc{\Kb}{\wbar{K}} \nc{\Kt}{\wtd{K}} \nc{\Kh}{\wht{K}}
\nc{\Lb}{\wbar{L}} \nc{\Lt}{\wtd{L}} \nc{\Lh}{\wht{L}}
\nc{\Mb}{\wbar{M}} \nc{\Mt}{\wtd{M}} \nc{\Mh}{\wht{M}}
\nc{\Nb}{\wbar{N}} \nc{\Nt}{\wtd{N}} \nc{\Nh}{\wht{N}}
\nc{\Ob}{\wbar{O}} \nc{\Ot}{\wtd{O}} \nc{\Oh}{\wht{O}}
\nc{\Pb}{\wbar{P}} \nc{\Pt}{\wtd{P}} \nc{\Ph}{\wht{P}}
\nc{\Qb}{\wbar{Q}} \nc{\Qt}{\wtd{Q}} \nc{\Qh}{\wht{Q}}
\nc{\Rb}{\wbar{R}} \nc{\Rt}{\wtd{R}} \nc{\Rh}{\wht{R}}
\nc{\Sb}{\wbar{S}} \nc{\St}{\wtd{S}} \nc{\Sh}{\wht{S}}
\nc{\Tb}{\wbar{T}} \nc{\Tt}{\wtd{T}} \nc{\Th}{\wht{T}}
\nc{\Ub}{\wbar{U}} \nc{\Ut}{\wtd{U}} \nc{\Uh}{\wht{U}}
\nc{\Vb}{\wbar{V}} \nc{\Vt}{\wtd{V}} \nc{\Vh}{\wht{V}}
\nc{\Wb}{\wbar{W}} \nc{\Wt}{\wtd{W}} \nc{\Wh}{\wht{W}}
\nc{\Xb}{\wbar{X}} \nc{\Xt}{\wtd{X}} \nc{\Xh}{\wht{X}}
\nc{\Yb}{\wbar{Y}} \nc{\Yt}{\wtd{Y}} \nc{\Yh}{\wht{Y}}
\nc{\Zb}{\wbar{Z}} \nc{\Zt}{\wtd{Z}} \nc{\Zh}{\wht{Z}}

\nc{\CA}{\mcl{A}} \nc{\CAb}{\wbar{\CA}} \nc{\CAt}{\wtd{\CA}} \nc{\CAh}{\wht{\CA}}
\nc{\CB}{\mcl{B}} \nc{\CBb}{\wbar{\CB}} \nc{\CBt}{\wtd{\CB}} \nc{\CBh}{\wht{\CB}}
\nc{\cD}{\mcl{D}} \nc{\cDb}{\wbar{\cD}} \nc{\cDt}{\wtd{\cC}} \nc{\cDh}{\wht{\cD}}
\nc{\CE}{\mcl{E}} \nc{\CEb}{\wbar{\CE}} \nc{\CEt}{\wtd{\CE}} \nc{\CEh}{\wht{\CE}}
\nc{\CF}{\mcl{F}} \nc{\CFb}{\wbar{\CF}} \nc{\CFt}{\wtd{\CF}} \nc{\CFh}{\wht{\CF}}
\nc{\CG}{\mcl{G}} \nc{\CGb}{\wbar{\CG}} \nc{\CGt}{\wtd{\CG}} \nc{\CGh}{\wht{\CG}}
\nc{\CH}{\mcl{H}} \nc{\CHb}{\wbar{\CH}} \nc{\CHt}{\wtd{\CH}} \nc{\CHh}{\wht{\CH}}
\nc{\CI}{\mcl{I}} \nc{\CIb}{\wbar{\CI}} \nc{\CIt}{\wtd{\CI}} \nc{\CIh}{\wht{\CI}}
\nc{\CJ}{\mcl{J}} \nc{\CJb}{\wbar{\CJ}} \nc{\CJt}{\wtd{\CJ}} \nc{\CJh}{\wht{\CJ}}
\nc{\CK}{\mcl{K}} \nc{\CKb}{\wbar{\CK}} \nc{\CKt}{\wtd{\CK}} \nc{\CKh}{\wht{\CK}}
\nc{\CL}{\mcl{L}} \nc{\CLb}{\wbar{\CL}} \nc{\CLt}{\wtd{\CL}} \nc{\CLh}{\wht{\CL}}
\nc{\CM}{\mcl{M}} \nc{\CMb}{\wbar{\CM}} \nc{\CMt}{\wtd{\CM}} \nc{\CMh}{\wht{\CM}}
\nc{\CN}{\mcl{N}} \nc{\CNb}{\wbar{\CN}} \nc{\CNt}{\wtd{\CN}} \nc{\CNh}{\wht{\CN}}
\nc{\CO}{\mcl{O}} \nc{\COb}{\wbar{\CO}} \nc{\COt}{\wtd{\CO}} \nc{\COh}{\wht{\CO}}
\nc{\CQ}{\mcl{Q}} \nc{\CQb}{\wbar{\CQ}} \nc{\CQt}{\wtd{\CQ}} \nc{\CQh}{\wht{\CQ}}
\nc{\CR}{\mcl{R}} \nc{\CRb}{\wbar{\CR}} \nc{\CRt}{\wtd{\CR}} \nc{\CRh}{\wht{\CR}}
\nc{\CS}{\mcl{S}} \nc{\CSb}{\wbar{\CS}} \nc{\CSt}{\wtd{\CS}} \nc{\CSh}{\wht{\CS}}
\nc{\CT}{\mcl{T}} \nc{\CTb}{\wbar{\CT}} \nc{\CTt}{\wtd{\CT}} \nc{\CTh}{\wht{\CT}}
\nc{\CU}{\mcl{U}} \nc{\CUb}{\wbar{\CU}} \nc{\CUt}{\wtd{\CU}} \nc{\CUh}{\wht{\CU}}
\nc{\CV}{\mcl{V}} \nc{\CVb}{\wbar{\CV}} \nc{\CVt}{\wtd{\CV}} \nc{\CVh}{\wht{\CV}}
\nc{\CW}{\mcl{W}} \nc{\CWb}{\wbar{\CW}} \nc{\CWt}{\wtd{\CW}} \nc{\CWh}{\wht{\CW}}
\nc{\CX}{\mcl{X}} \nc{\CXb}{\wbar{\CX}} \nc{\CXt}{\wtd{\CX}} \nc{\CXh}{\wht{\CX}}
\nc{\CY}{\mcl{Y}} \nc{\CYb}{\wbar{\CY}} \nc{\CYt}{\wtd{\CY}} \nc{\CYh}{\wht{\CY}}
\nc{\CZ}{\mcl{Z}} \nc{\CZb}{\wbar{\CZ}} \nc{\CZt}{\wtd{\CZ}} \nc{\CZh}{\wht{\CZ}}

\let\eps\epsilon
\let\ups\upsilon

\let\veps\varepsilon
\let\vtht\vartheta
\let\vsgm\varsigma
\let\vphi\varphi
\let\vrho\varrho

\nc{\alphab}{\bar{\alpha}} \nc{\alphat}{\td{\alpha}} \nc{\alphah}{\hat{\alpha}}
\nc{\betab}{\bar{\beta}}   \nc{\betat}{\td{\beta}}   \nc{\betah}{\hat{\beta}} 
\nc{\gammab}{\bar{\gamma}} \nc{\gammat}{\td{\gamma}} \nc{\gammah}{\hat{\gamma}} 
\nc{\deltab}{\bar{\delta}} \nc{\deltat}{\td{\delta}} \nc{\deltah}{\hat{\delta}} 
\nc{\epsilonb}{\bar{\eps}} \nc{\epsilont}{\td{\eps}} \nc{\epsilonh}{\hat{\eps}} 
\nc{\vepsb}{\bar{\veps}}   \nc{\vepst}{\td{\veps}}   \nc{\vepsh}{\hat{\veps}} 
\nc{\zetab}{\bar{\zeta}}   \nc{\zetat}{\td{\zeta}}   \nc{\zetah}{\hat{\zeta}} 
\nc{\etab}{\bar{\eta}}     \nc{\etat}{\td{\eta}}     \nc{\etah}{\hat{\eta}} 
\nc{\thetab}{\bar{\theta}} \nc{\thetat}{\td{\theta}} \nc{\thetah}{\hat{\theta}} 
\nc{\vthetab}{\bar{\vtht}} \nc{\vthetat}{\td{\vtht}} \nc{\vthetah}{\hat{\vtht}} 
\nc{\lambdab}{\bar{\lambda}} \nc{\lambdat}{\td{\lambda}} \nc{\lambdah}{\hat{\lambda}} 
\nc{\iotab}{\bar{\iota}}   \nc{\iotat}{\td{\iota}}   \nc{\iotah}{\hat{\iota}} 
\nc{\kappab}{\bar{\kappa}} \nc{\kappat}{\td{\kappa}} \nc{\kappah}{\hat{\kappa}} 
\nc{\lmdb}{\bar{\lmd}}     \nc{\lmdt}{\td{\lmd}}     \nc{\lmdh}{\hat{\lmd}} 
\nc{\mub}{\bar{\mu}}       \nc{\mut}{\td{\mu}}       \nc{\muh}{\hat{\mu}} 
\nc{\nub}{\bar{\nu}}       \nc{\nut}{\td{\nu}}       \nc{\nuh}{\hat{\nu}} 
\nc{\xib}{\bar{\xi}}       \nc{\xit}{\td{\xi}}       \nc{\xih}{\hat{\xi}} 
\nc{\pib}{\bar{\pi}}       \nc{\pit}{\td{\pi}}       \nc{\pih}{\hat{\pi}} 
\nc{\vpib}{\bar{\vpi}}     \nc{\vpit}{\td{\vpi}}     \nc{\vpih}{\hat{\vpi}} 
\nc{\rhob}{\bar{\rho}}     \nc{\rhot}{\td{\rho}}     \nc{\rhoh}{\hat{\rho}} 
\nc{\vrhob}{\bar{\vrho}}   \nc{\vrhot}{\td{\vrho}}   \nc{\vrhoh}{\hat{\vrho}} 
\nc{\sigmab}{\bar{\sigma}} \nc{\sigmat}{\td{\sigma}} \nc{\sigmah}{\hat{\sigma}} 
\nc{\vsigmab}{\bar{\vsgm}} \nc{\vsigmat}{\td{\vsgm}} \nc{\vsigmah}{\hat{\vsgm}} 
\nc{\taub}{\bar{\tau}}     \nc{\taut}{\td{\tau}}     \nc{\tauh}{\hat{\tau}} 
\nc{\upsb}{\bar{\ups}} \nc{\upst}{\td{\ups}} \nc{\upsh}{\hat{\ups}} 
\nc{\phib}{\bar{\phi}}     \nc{\phit}{\td{\phi}}     \nc{\phih}{\hat{\phi}} 
\nc{\varphib}{\bar{\vphi}}   \nc{\varphit}{\td{\vphi}}   \nc{\varphih}{\hat{\vphi}} 
\nc{\chib}{\bar{\chi}}     \nc{\chit}{\td{\chi}}     \nc{\chih}{\hat{\chi}} 
\nc{\omegab}{\bar{\omega}} \nc{\omegat}{\td{\omega}} \nc{\omegah}{\hat{\omega}} 

\nc{\Gammab}{\wbar{\Gamma}}     \nc{\Gammat}{\wtd{\Gamma}}     \nc{\Gammah}{\wht{\Gamma}}
\nc{\Deltab}{\wbar{\Delta}}     \nc{\Deltat}{\wtd{\Delta}}     \nc{\Deltah}{\wht{\Delta}}
\nc{\Thetab}{\wbar{\Theta}}     \nc{\Thetat}{\wtd{\Theta}}     \nc{\Thetah}{\wht{\Theta}}
\nc{\Lambdab}{\wbar{\Lambda}}   \nc{\Lambdat}{\wtd{\Lambda}}   \nc{\Lambdah}{\wht{\Lambda}}
\nc{\Xib}{\wbar{\Xi}}           \nc{\Xit}{\wtd{\Xi}}           \nc{\Xih}{\wht{\Xi}}
\nc{\Pib}{\wbar{\Pi}}           \nc{\Pit}{\wtd{\Pi}}           \nc{\Pih}{\wht{\Pi}}
\nc{\Sigmab}{\wbar{\Sigma}}     \nc{\Sigmat}{\wtd{\Sigma}}     \nc{\Sigmah}{\wht{\Sigma}}
\nc{\Upsilonb}{\wbar{\Upsilon}} \nc{\Upsilont}{\wtd{\Upsilon}} \nc{\Upsilonh}{\wht{\Upsilon}}
\nc{\Phib}{\wbar{\Phi}}         \nc{\Phit}{\wtd{\Phi}}         \nc{\Phih}{\wht{\Phi}}
\nc{\Psib}{\wbar{\Psi}}         \nc{\Psit}{\wtd{\Psi}}         \nc{\Psih}{\wht{\Psi}}
\nc{\Omegab}{\wbar{\Omega}}     \nc{\Omegat}{\wtd{\Omega}}     \nc{\Omegah}{\wht{\Omega}}

\nc{\txd}{d}

\newcommand{\pa}{\partial}
\nc{\tcos}{\textrm{cos}}
\nc{\tsin}{\textrm{sin}}

\nc{\al}{\alpha}
\nc{\bt}{\beta}
\nc{\gm}{\gamma}
\nc{\rh}{\rho}
\nc{\zt}{\zeta}
\nc{\Dl}{\Delta}
\nc{\TD}{\til{\Delta}}

\nc{\sg}{\Sigma}

\nc{\rd}{0.75}

\def\ie{\begin{equation}\begin{aligned}}
\def\fe{\end{aligned}\end{equation}}
\begin{document}

\title{R-matrices from Feynman Diagrams in 5d Chern-Simons Theory and Twisted M-theory }
\author{Meer Ashwinkumar}
\affiliation{Albert Einstein Center for Fundamental Physics, Institute for Theoretical Physics,
University of Bern, Sidlerstrasse 5, CH-3012 Bern, Switzerland}
\emailAdd{meer.ashwinkumar@unibe.ch}
\abstract{
In this work, we study the analogues of R-matrices that arise in 5d non-commutative topological-holomorphic Chern-Simons theory, which is known to describe twisted M-theory. We first study the intersections of line and surface operators in 5d Chern-Simons theory, which correspond to M2- and M5-branes, respectively. A Feynman diagram computation of  the correlation function of this configuration furnishes an expression reminiscent of an R-matrix derivable from 4d Chern-Simons theory.
We explain how this object is related to a Miura operator that is known to realize (matrix-extended) $W_{\infty}$-algebras. For 5d Chern-Simons theory with nonabelian gauge group, we then perform a Feynman diagram computation of coproducts for deformed double current algebras and matrix-extended $W_{\infty}$-algebras from fusions of M2-branes, M5-branes, and M2-M5 intersections. }

\maketitle

\noindent
\section{Introduction}

Five-dimensional non-commutative topological-holomorphic Chern-Simons theory was introduced by Costello in \cite{Costello:2016nkh} as an approach to describing twisted M-theory, also known as $\Omega$-deformed M-theory, on a  Taub-NUT manifold with $A_{K-1}$ singularity at the origin. This Chern-Simons theory is of the form
\begin{equation}\label{act1}
-\frac{1}{4 \pi \epsilon_1 } \int_{\mathbb{R}\times \mathbb{C}\times \mathbb{C}}  d w \wedge d z \wedge \operatorname{Tr} A \wedge *_{\epsilon_2} d A-\frac{1}{6 \pi \epsilon_1} \int_{\mathbb{R}\times \mathbb{C}\times \mathbb{C}}  d w \wedge d z \wedge  \operatorname{Tr} A \wedge *_{\epsilon_2} A \wedge *_{\epsilon_2} A 
\end{equation}
where $*_{\epsilon_2 }$ is the Moyal product that depends on a parameter $\epsilon_2$, defined as
\begin{equation}
f * g=f g+\epsilon_2 \frac{1}{2} \varepsilon_{i j} \frac{\partial}{\partial z_i} f \frac{\partial}{\partial z_j} g+\epsilon_2^2 \frac{1}{2^2 \cdot 2!} \varepsilon_{i_{1 j} j_1} \varepsilon_{i j_2}\left(\frac{\partial}{\partial z_{i_1}} \frac{\partial}{\partial z_{i_2}} f\right)\left(\frac{\partial}{\partial z_{j_1}} \frac{\partial}{\partial z_{j_2}} g\right)+\ldots
\end{equation}
for elements $f,g$ of the ring of holomorphic functions on $\mathbb{C}_w \times \mathbb{C}_z$, where $z_1=w$ and $z_2=z$, where $\varepsilon_{ij}$ is the alternating symbol and the summation convention has been utilized.
The Moyal product arises from the non-commutativity of the holomorphic coordinates $z$ and $w$ of the two complex planes on which the theory is defined. In addition, $\epsilon_1$ is the coupling constant of the theory, which we shall also denote interchangeably as $\hbar$.\footnote{We shall not introduce a third $\Omega$-deformation parameter denoted $\epsilon_3$ as done in some references, such as  \cite{Gaiotto:2019wcc,Oh:2020hph}, and shall instead follow the conventions of \cite{Costello:2016nkh}.} The theory is holomorphic along the two complex planes, but is otherwise topological along the remaining direction $\mathbb{R}$, which is parameterized by the coordinate $t$. Here, $A$ is a partial connection of the form 
\begin{equation}
A=A_tdt  + A_{\wb} d\wb  +A_{\zb} d\zb.
\end{equation}

As explained in \cite{Costello:2016nkh}, the aforementioned twist of M-theory is insensitive to the radius of the 11th circle of M-theory, which is taken to be the circle fiber of the Taub-NUT manifold, implying that the twisted theory is equivalent to type IIA string theory with a stack of $K$ D6-branes in a $B$-field background, with $\Omega$-deformation along two dimensions. The latter is responsible for the localization of the D6-brane worldvolume theory to the 5d non-commutative Chern-Simons theory \eqref{act1} with gauge group $GL(K)$, in analogy to how 4d and 3d Chern-Simons theories arise from the D5- and D4-brane worldvolume theory subject to an $\Omega$-deformation, respectively \cite{Luo:2014sva,Costello:2018txb}.

From a mathematical perspective, this Chern-Simons theory admits line operators in representations of a quantum algebra related to the affine Yangian, and significant progress has been made in interpreting the Koszul dual of this algebra in terms of twisted holography for 5d Chern-Simons theory, see, e.g., \cite{Costello:2017fbo,Oh:2020hph,Oh:2021bwi,Oh:2021wes}.
Notably, M2 and M5-branes can be identified with line operators and surface operators, respectively, in 5d non-commutative Chern-Simons theory.

In this work, we shall study correlation functions of  intersections of these line and surface operators in 5d Chern-Simons theory, describing intersecting M2- and M5-branes in twisted M-theory in the process. 
As we shall observe, the correlation function is similar in form to the quasi-classical expansion of a rational, spectral parameter-dependent R-matrix. 
This quantity is derivable from intersecting line defects in 4d Chern-Simons theory \cite{Costello:2013zra,Costello:2017dso},
which has the form 
\begin{equation}\label{4dCS}
S=\frac{i}{2\pi\hbar}\int_{ \mathbb{R}^2\times \mathbb{C}} dz \wedge \textrm{Tr}\bigg(A\wedge d A + \frac{2}{3}  A\wedge  A\wedge  A\bigg).
\end{equation} 
In the latter theory, the correlation function of two line operators at points $z_1$ and $z_2$ on $\C$, and in representations $R_1$ and $R_2$ of the generators of $G$, realizes the $R$-matrix
\begin{equation}
\til{R}_{12}(z_1-z_2)=\mathds{1}+\frac{\hbar}{z_1-z_2}T_{R_1}^a\otimes T_{ R_2 a}+{O}(\hbar^2)
\end{equation}
to linear order in $\hbar$. As we shall see, we obtain from 5d Chern-Simons theory a generalization of this expression that reflects the additional holomorphic dependence of the theory on the $w$-plane.

This result follows naturally, given that the 4d and 5d (commutative) Chern-Simons theories are related via a field-theoretic T-duality \cite{Yamazaki:2019prm}. A straightforward approach to understanding this relationship is by noticing that if we replace one of the complex planes in the action \eqref{act1} with the cylinder $S^1 \times \mathbb{R}$, take the commutative limit $\epsilon_2 \rightarrow 0$, and dimensionally reduce along the circle direction, one ends up with the 4d Chern-Simons theory action \eqref{4dCS}.  

We shall now provide a brief overview of this work.
In Section \ref{sec2}, we derive the aforementioned R-matrix-like operator in 5d Chern-Simons theory. We shall explain how to relate this operator to an elementary Miura operator for (matrix-extended) $W_{\infty}$ algebras, providing evidence for a proposal of Gaiotto and Rapcak \cite{Gaiotto:2020dsq}. We shall further interpret the realization of the general Miura transformation (which furnishes the $W$-algebras of interest) in terms of the concatenation of multiple M5-branes intersecting a single M2-brane represented by a Wilson line.  

The Miura operator is expected to play the role of intertwiner for representations of the deformed double current algebra (associated with M2-branes/Wilson lines) and (matrix-extended) $W_{\infty}$ algebras (associated with M5-branes/surface operators).  In Section \ref{sec3}, we shall perform Feynman diagram computations of various coproducts for the deformed double current algebra and the matrix-extended $W_{\infty}$ algebra from the fusion of M2-branes, the fusion of M5-branes, and gauge-invariance constraints of M2-M5 intersections. These coproducts were previously derived through alternate, rigorous methods in \cite{Gaiotto:2023ynn}. The coproducts are then related to the Yang-Baxter equation that the Miura operator solves. 
\newline
\newline
\noindent \textbf{Note :} As this paper was nearing completion, the author was made aware of work by N. Ishtiaque, S. Jeong, and Y. Zhou \cite{Ishtiaque:2024orn}, with overlapping content. The author would like to state that the results contained in this work were obtained independently. The author would also like to thank the aforementioned researchers for coordinating the submission of our papers to the arXiv.

\acknowledgments
 	
 	We would like to thank Roland Bittleston, Matthias Blau, Shigenori Nakatsuka,
 	Meng-Chwan Tan, Junya Yagi and  Masahito Yamazaki for helpful discussions. We would also like to thank an anonymous referee for useful feedback on a previous version of this manuscript.
Initial work on this paper was performed at Kavli IPMU, where the author was supported by the JSPS Grants-in-Aid for Scientific Research No.
19H00689 and 20H05860.

\section{M2-M5 Intersections}\label{sec2}

To compute the interaction between an ordinary Wilson line and a holomorphic Wilson line, we first require the propagator of 5d Chern-Simons theory. Although this propagator has been presented before in multiple references \cite{Costello:2017fbo,Oh:2020hph}, we shall derive it explicitly  while being careful to fix its normalization, which is crucial 
for our applications.
This derivation can be performed in a manner similar to the computation of the propagator in 4d Chern-Simons theory by using  the analogue of the Lorentz  gauge \cite{Costello:2017dso}.

Before proceeding, let us provide motivation for the choice of Lorentz gauge. The use of the Lorentz gauge in Chern--Simons theories for analyzing the convergence and well-definedness of Feynman diagrams goes back to the work of Bott--Taubes \cite{bott1994self} and Axelrod--Singer \cite{Axelrod:1991vq, axelrod1994chern}. However,  common alternate gauges are the light-cone gauge where a light-cone component of the gauge field is set to zero \cite{cmp/1104179728}, or the axial gauge where the time component of the gauge field is set to zero  \cite{Cattaneo:2020lle}. Gauges of the latter type are however inadequate for the analysis of quantum interactions involving the 3-point interaction vertex of a Chern--Simons theory, since this vertex vanishes in such gauges. This motivates us to choose the Lorentz gauge as it will be necessary for computations in Section \ref{sec3}, which involve evaluating Feynman diagrams involving interaction vertices. However,  we shall also use an analogue of the lightcone  gauge (known as the holomorphic gauge) when performing a check of our computation of the line-surface interaction in Section \ref{2.2}. 

The Lorentz gauge in the present case of 5d Chern-Simons theory takes the form 
\ie \label{gf}
\partial_t A_t + 4 \partial_w A_{\bar{w}}+ 4 \partial_z A_{\bar{z}}=0.
\fe
The factors of 4 are due to the fact that, with the choice of metric
\ie 
ds^2=dt^2+ dw d\bar{w}+ dz d\bar{z}
\fe
on $\mathbb{R}\times \mathbb{C} \times \mathbb{C}$, the gauge-fixing condition \eqref{gf} and the linearized equations of motion $dw\wedge dz \wedge dA=0$ together imply that each gauge field component is harmonic, i.e., they satisfy
\ie 
g^{\mu \nu} \frac{\partial}{\partial x^\mu} \frac{\partial}{\partial x^\nu}A_i=\left(\frac{\partial^2}{\partial x^2}+4\frac{\partial}{\partial w}\frac{\partial}{\partial \bar{w}}+4 \frac{\partial}{\partial z} \frac{\partial}{\partial \bar{z}}\right) A_i=0,
\fe
where $i=t,\bar{w},\bar{z}$, and where $g^{\mu \nu}$ has the components $g^{tt}=1$, $g^{w\bar{w}}=2$ and $g^{z\bar{z}}=2$.  In addition, we impose a boundary condition on the gauge field that sets it to zero at infinity on each of the complex planes $\mathbb{C}_w$ and $\mathbb{C}_z$ as well as at the endpoints of $\mathbb{R}_t$, and the propagator ought to satisfy this boundary condition.
 
It shall prove convenient to express the propagator as a two-form on two copies of $\R\times \C \times \C $. The following propagator two-form with adjoint indices removed shall also be employed:
\ie 
P^{a b}(t, w, \bar{w}, z, \bar{z})=\delta^{a b} P(t, w, \bar{w},z, \bar{z}).
\fe
Now, the defining equations of this  propagator  
two-form are 
\begin{align}
-\frac{\mathrm{1}}{4 \pi} dw \wedge{d} z \wedge  \mathrm{d} P(t,w, \bar{w}, z, \bar{z}) & =\delta_{t,w,\bar{w}, z, \bar{z}=0} \label{propeq1} \\
\left(\partial_t \iota_{t}+4 \partial_w \iota_{\bar{w}}+4 \partial_z \iota_{\bar{z}}\right) P(t, w, \bar{w},z, \bar{z}) & =0 . \label{propeq2}
\end{align}
where $\delta_{t,w, \bar{w}, z, \bar{z}=0} $ denotes a delta-function distribution 5-form. The second defining equation imposes the analogue of the Lorentz gauge at the level of the propagator.  
The claim is that, for
$t=t'-t''$, $w=w'-w''$, and $z=z'-z''$, the propagator two-form is  
\begin{equation}
    \begin{aligned}\label{eq.propagator}
P^{ab}(t,z, \bar{z},w, 
\bar{w})&:=
\frac{1}{2}  \langle A^a_i (t', z', \bar{z}', w', \bar{w}') A^b_j(t'', z'', \bar{z}'', w'', \bar{w}'')\rangle d x^i \wedge d x^j 
\\
&= \frac{3}{4}\frac{\delta^{ab}}{2 \pi}\left( t d 
\wb \wedge d \zb + 2 \wb d \zb \wedge d t +2 \bar{z}
d t \wedge d \wb \right)\frac{1}{(t^2+w \bar{w}+z \bar{z})^{5/2}} 
 \;.
\end{aligned}
\end{equation}

Let us verify that this is indeed the correct propagator, verifying its normalization along the way. Away from the origin, the RHS of \eqref{propeq1} ought to be zero. We can check that $dw \wedge dz \wedge dP $ is indeed equal to
 \ie 
dw\wedge dz \wedge \bigg[& \frac{3}{4}\frac{1}{2\pi} \frac{5 dt \wedge d\bar{w} \wedge d\bar{z}}{(t^2+|w|^2+|z|^2)^{5/2}}\\&-\frac{3}{4}\frac{1}{2\pi}\frac{5}{2} \frac{(2t dt +\bar{w} dw + w d \bar{w}+\bar{z} dz + z d \bar{z})}{(t^2+|w|^2+|z|^2)^{7/2}} \wedge(t d 
\wb \wedge d \zb + 2 \wb d \zb \wedge d t +2 \bar{z}
d t \wedge d \wb)\bigg]\\=0.
\fe 
Moreover, we can verify \eqref{propeq2}, i.e., that 
$\left(\partial_t \iota_{t}+4 \partial_w \iota_{{\bar{w}}}+4 \partial_z \iota_{\bar{z}}\right) P(t, w,\bar{w}, z, \bar{z})$ is proportional to 
\ie 
&\partial_t \left(\frac{-2\bar{w} d\bar{z} +2 \bar{z}d\bar{w} }{(t^2+|w|^2+|z|^2)^{5/2}} \right) + 4\partial_{{z}}\left(\frac{-t d\bar{w} +2 \bar{w}dt }{(t^2+|w|^2+|z|^2)^{5/2}} \right) +4\partial_{{w}}\left(\frac{t d\bar{z} -2 \bar{z} dt }{(t^2+|w|^2+|z|^2)^{5/2}} \right) \\
& = \frac{5}{2}\left( \frac{(-4t\bar{w} d\bar{z} +4 t\bar{z} d\bar{w}  )}{(t^2+|w|^2+|z|^2)^{7/2}} +4 \frac{(-t \bar{z} d\bar{w} + 2  \bar{w}\bar{z} dt )}{(t^2+|w|^2+|z|^2)^{7/2}} +4 \frac{(\bar{w} t d\bar{z} - 2 \bar{w}\bar{z} dt )}{(t^2+|w|^2+|z|^2)^{7/2}}\right)\\&=0.
\fe 
The normalization of the propagator is fixed by checking \eqref{propeq1} at the origin. The first step in this direction is the observation that the propagator two-form restricted to a four-sphere of radius $\epsilon> 0$ takes the form 
\ie \label{theeq}
P(t, w, \bar{w}, z, \bar{z})=\frac{3}{4}\frac{1}{2 \pi}\frac{1}{\epsilon^5}(t d 
\wb \wedge d \zb + 2 \wb d \zb \wedge d t +2 \bar{z}
d t \wedge d \wb).
\fe
If we were to integrate the propagator over a five-ball of radius $\epsilon$ (which has volume $8\pi^2\epsilon^5/15$), then Stokes' theorem tells us that we can use \eqref{theeq} in the computation, i.e., 
\ie 
\begin{aligned}
-\frac{1}{4 \pi} \int_{t^2+w \bar{w}+z \bar{z} \leq \epsilon} \mathrm{~d} w \wedge dz\wedge \mathrm{d} P(t, w,\bar{w}, z, \bar{z}) & =-\frac{3}{4}\frac{1}{4 \pi} \frac{1}{2 \pi} \frac{1}{\epsilon^5}\int_{t^2+w \bar{w}+z \bar{z}\leq \epsilon} 5 \mathrm{~d} t \mathrm{~d} w \mathrm{~d} \bar{w} \mathrm{~d} z \mathrm{~d} \bar{z} \\
& =-\frac{3}{4}\frac{1}{4 \pi} \frac{1}{2 \pi} \frac{1}{\epsilon^5}\int_{t^2+w \bar{w}+z \bar{z} \leq \epsilon}5(-4 ) \mathrm{d} t \mathrm{~d} u \mathrm{~d} v \mathrm{~d} u' \mathrm{~d} v' \\
& =\frac{3}{4}\frac{1}{4 \pi} \frac{1}{2 \pi}(5)(4 ) \frac{8\pi^2}{15}=1, 
\end{aligned}
\fe
where $w=u + i v$ and $z=u' + i v'$. 

Finally, we note that the propagator has the correct behavior at infinity on $\mathbb{R} \times \mathbb{C} \times \mathbb{C} $. This completes the verification of the form and normalization of the propagator 
\eqref{eq.propagator}.

\subsection{M2-branes as Wilson Lines and M5-branes as Surface Operators}

M2- and M5-branes can be described respectively as Wilson lines and surface operators in 5d Chern-Simons (CS) theory \cite{Costello:2016nkh}.

To be precise, in the reduction of twisted M-theory on the Taub-NUT fiber that forms the 11-th circle of M-theory to type IIA string theory, one considers M2-branes that are transverse to the 11-th circle, that reduce to D2-branes. The remaining $\Omega$-background localizes the worldvolume theory of these D2-branes to a quantum mechanical system that quantizes to a Wilson line along the topological direction parameterized by $t$. 

The M5-branes of interest wrap the 11-th circle, and reduce to D4-branes intersecting the $K$ D6-branes along the complex surface parameterized by $w,\bar{w}$. For a stack of $N$ M5-branes, this system can be described by $NK$ free chiral fermions \cite{Dijkgraaf:2007sw,Costello:2016nkh}.

Wilson lines in 5d CS can be defined in a familiar manner as a path ordered exponential of the integral of the gauge field (in a fixed representation) over the $\mathbb{R}$ direction, explicitly given as 
\ie \label{po}
&P e^{\int_{-\infty}^{\infty} A_{t}dt }\\=& \mathds{1} + \int_{-\infty}^{\infty} dt A_{t}(t)  + \int_{-\infty}^{\infty} dt 'A_{t'} (t') \int_{-\infty}^{t'}dt A_{t} (t)\\&+\int_{-\infty}^{\infty} dt''   A_{t''} (t'') \int_{-\infty}^{t''}   dt' A_{t'} (t') \int_{-\infty}^{t'} dt A_{t} (t)+\ldots 
\fe 
We shall also consider Wilson lines associated with representations of the double polynomial loop algebra $\mathfrak{g}[[z,w]]$, where the coupling to the gauge field is 
\ie \label{wilson}
\frac{1}{m!}\frac{1}{n!}\partial_z^m \partial_w^n A^a t_a[m,n].
\fe
In analogy to the case of 4d Chern-Simons theory \cite{Costello:2017dso}, where Wilson lines in representations of $\mathfrak{g}[[z]]$ arise naturally by taking OPEs of ordinary Wilson lines, such OPEs in 5d Chern-Simons lead us naturally to Wilson lines in representations of 
$\mathfrak{g}[[z,w]]$ (we shall show this explicitly in the case of nonabelian gauge group). In fact, as shown in \cite{Costello:2017fbo}, a 1-loop quantum correction to the condition for gauge invariance of such a Wilson line implies that it is actually associated with a deformation of $\mathfrak{g}[[z,w]]$ known as the deformed double current algebra, which is related to the affine Yangian. This is in analogy to the fact that (non-anomalous) Wilson lines in 4d Chern-Simons theory are in representations of the Yangian at the quantum level \cite{Costello:2017dso}. 

The holomorphic surface operator supporting chiral free fermions and describing M5-branes has an action of the form
\begin{equation}\label{ff}
\int_{\mathbb{C}_w} d^2 w \operatorname{Tr} \psi(\partial_{\bar{w}}+A_{\bar{w}}) \psi^{\prime}.
\end{equation}
Here, the free fermions transform in the fundamental representation with respect to the $GL(K)$ gauge group as well as the $GL(N)$ flavor symmetry. This surface operator suffers from a gauge anomaly, 
that can be cancelled by
including a coupling to a Beltrami differential as in \cite{Costello:2020jbh},\footnote{To be precise, the  coupling to the Beltrami differential for $N$ M5-branes wrapping the $w$-plane would be proportional to $ \int_{\mathbb{R}\times \mathbb{C}^2}A^a\mu_{BR}A_a dz\wedge dw$, where $\mu_{BR}$ is the Beltrami differential. Examples of choices for $\mu_{BR}$ (for surface operators at $t=0$ and $z=0$) are  $\mu_{BR}=N\frac{2\bar{w}dt - t d\bar{z}}{(t^2+|z|^2)^{\frac{3}{2}}}\partial_w$ and $\mu_{BR}=N \theta(t)\delta(z,\bar{z}) d\bar{z} \partial_w$, where $\theta(t)$ is the step function. In addition, gauge transformation of the gauge field needs to be modified appropriately, using a mechanism analogous to that of \cite{Costello:2020lpi}.} which can be interpreted as the backreaction of the corresponding M5-branes. We also expect that the anomaly can be cancelled by deforming the geometry of the directions transverse to $\mathbb{C}_w$ as in \cite{Khan:2022vrx}, although this has not been shown explicitly. The coefficient of the gauge field in \eqref{ff} is a current that generates a $\widehat{\mathfrak{gl}}(K)$ affine Kac-Moody algebra, which we shall describe in further detail below.  We also note that we shall use the convention $d^2w = \frac{dw d \bar{w}}{-i2}$ for the measure of these surface operators.

More generally, as in \eqref{wilson}, we can consider a surface defect with holomorphic $z$-derivatives of the gauge fields coupled to currents of higher spin
\begin{equation}\label{gensurf}
\int_{\mathbb{C}_w} d^2w W^{(m)} \partial_z^{m-1} A, 
\end{equation}
where $m$ takes value in $1,\ldots, p$ for some positive integer $p$.
In \cite{Costello:2016nkh}, a mathematically rigorous approach to taking the large $p$ limit was proposed by Costello, and the M5-brane algebra was identified with the $W_{\infty}$-algebra. One of our aims is to explicitly show that the elementary Miura operators that realize $W_{\infty}$-algebras can be derived using Feynman diagrams in 5d Chern-Simons theory. 

As mentioned above, the currents of the form $J=\psi \psi'$ that enter \eqref{ff} generate an affine Kac-Moody algebra. For the case of a single M5-brane ($N=1$) and $K=1$, corresponding to $GL(1)$ 5d Chern-Simons theory, the OPE of the currents is\footnote{The shift of the level by a term proportional $\epsilon_2$ can be understood by studying the corrections to the gauge anomaly of the surface defect due to the non-commutativity of $\mathbb{C}^2$ \cite{Costello:2016nkh}.  } 
\ie \label{u1ca}
J(w) J(w') \sim \frac{\epsilon_1+ \epsilon_2}{(w-w')^2}.
\fe
For $N=1$, but $K$ arbitrary, the surface operator for $GL(K)$ 5d Chern-Simons theory supports affine Kac-Moody currents with the OPE 
\begin{equation}
J^{ a}{ }_b(w) J^{ c}{ }_d(w') \sim \frac{\epsilon_1\delta^a{ }_b \delta^c{ }_d+\epsilon_2 \delta^c{ }_b \delta^a{ }_d}{(w-w')^2}+\frac{\delta^c{ }_b J^{ a}{ }_d(w')-\delta^a{ }_d J^{ c}{ }_b(w')}{w-w'}.
\end{equation}

It is worth noting that for 6d holomorphic Chern-Simons theory, Costello and Paquette \cite{Costello:2020jbh} defined a general form for holomorphic surface operators that is reminiscent of the path ordering of the Wilson line in \eqref{po}:
\ie 
\sum_{n \geq 0} \frac{1}{n!} \int_{z_1, \ldots, z_n \in \mathbb{C}} \prod_{i=1}^n\left(\int \frac{1}{k_1^{i}!k_2^{i}!} \partial_{w_1}^{k_1^i} \partial_{w_2}^{k_2^i} A_{\bar{z}}^{a_i}\left(z_i\right) J_{a_i}\left[k_1, k_2\right]\left(z_i\right)\right),
\fe 
where $z$, $w_1$ and $w_2$ are complex coordinates on which the 6d theory is defined, and where the currents must satisfy the OPE 
\ie 
J_b\left[l_1, l_2\right](0) J_c\left[m_1, m_2\right](z) \sim \frac{1}{z} f_{b c}^a J_a\left[l_1+m_1, l_2+m_2\right].
\fe 
for classical gauge invariance; quantum gauge invariance requires further corrections to this OPE. For the present case of 5d Chern-Simons theory, which has only two holomorphic directions, the surface operator would have the form
\ie 
\sum_{n \geq 0} \frac{1}{n!} \int_{w_1, \ldots, w_n \in \mathbb{C}} \prod_{i=1}^n\left(\int \frac{1}{m!} \partial_{z}^{m^i}  A_{\bar{w}}^{a_i}\left(w_i\right) J_{a_i}\left[ m\right]\left(w_i\right)\right).
\fe

\subsection{M2-M5 Intersection at Leading Order in $\hbar$}\label{2.2}

In this subsection, we shall perform an explicit derivation  of an analogue of an R-matrix known as the elementary Miura operator for matrix-extended $W_{\infty}$ algebras. Let us explain the reason that this configuration is expected to realize an R-matrix-like object. As explained above, in previous work on 5d Chern--Simons theory \cite{Costello:2016nkh,Costello:2017fbo,Oh:2021wes}, it was shown that line and surface defects in 5d Chern--Simons theory were associated with variants of the affine Yangian and $W_{\infty}$ algebras, respectively. The elementary Miura operators that can be used to derive $W_{\infty}$-algebras are known to satisfy a Yang--Baxter equation \cite{ Prochazka:2019dvu}, and therefore we expect that Miura operators can be derived from intersections of defects in 5d Chern--Simons theory, by analogy with the derivation of R-matrices for the Yangian algebra from correlation functions of crossed Wilson lines in \cite{Costello:2017dso}. Moreover, in \cite{Gaiotto:2020dsq}, it was conjectured that line-surface intersections ought to give rise to Miura operators in 5d Chern--Simons theory. 

\begin{figure}
\begin{center}
\begin{tikzpicture}[tdplot_main_coords]

  \fill[gray!30] (-2,-2,0) -- (2,-2,0) -- (2,2,0) -- (-2,2,0) -- cycle;
  \draw[thick] (-2,-2,0) -- (2,-2,0) -- (2,2,0) -- (-2,2,0) -- cycle;

  \draw[thick, blue] (0,0,-2) -- (0,0,2);

  \fill[red] (0,0,0) circle (1pt);
  
  \node[below right] at (2,2,0) {M5-brane};
  \node[right] at (0,0,2) {M2-brane};
  \node[below left] at (0,0,0) {};

\end{tikzpicture}
\end{center}
\caption{M2-M5-brane intersection as the intersection of a line and surface defect in 5d Chern-Simons theory.}
\label{fig1}
\end{figure}
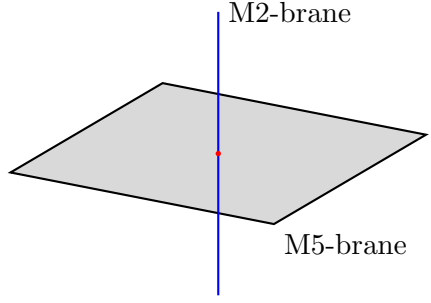

We shall verify this in what follows. 
As in \cite{Costello:2017dso}, we shall consider perturbation theory around the trivial classical solution $A=0$ when computing the correlation function of intersecting line and surface defects, depicted in Figure \ref{fig1}. We shall consider the leading nontrivial contribution to this correlation function.

\tdplotsetmaincoords{70}{120}

The order $\hbar$ coupling between an ordinary Wilson line, associated with a representation $R$ of $G$, and a holomorphic surface defect, arises from the correlation function 
\ie \label{fund}
&\langle\int_{-\infty}^{\infty} dt' \int d^2w''      A_{\wb}(t'' ,z'' ,w'' )A_t(t',z',w')\rangle\\=& -\hbar \int_{-\infty}^{\infty} dt' \int d^2w''   \left( \frac{3}{4}\frac{1}{2\pi}\right)\bigg(\frac{2
(\bar{z}'-\bar{z}'')}{((t'-t'')^2+|w'-w''|^2+|z'-z''|^2)^{5/2}}\bigg) J_{a}(w'')T_{R}^a .
\fe
To evaluate the integrals, we first Taylor expand the current $J_a(w'')$ as 
\ie 
J_a(w'')=\sum_{n\in \{0\} \cup \mathbb{N}} J_{an} (w'')^n.
\fe
The expression of interest can then be rewritten, after the shifts $t'\rightarrow t'+t"$ and $w'' \rightarrow w'' +w '$, as   
\ie 
 &-\hbar \int_{-\infty}^{\infty} dt' \int d^2 w'' \left( \frac{3}{4}\frac{1}{2\pi}\right) \bigg(\frac{2
(\bar{z}'-\bar{z}'')}{((t')^2+|w''|^2+|z'-z''|^2)^{5/2}}\bigg) \bigg(\sum_{n\in \{0\} \cup \mathbb{N}}  J_{an} (w''+w')^n\bigg)T_{R}^a .
\fe 
Performing a binomial expansion of the factor $(w''+w')^n$  to obtain 
$(w''+w')^n
= \sum_{k=0}^{n} \binom{n}{k} (w'')^{k} (w')^{\,n-k}$, and using polar coordinates on the $w''$-plane by setting $w''=r_{w''}e^{i\theta_{w''}}$, we find that only the factor of $(w')^n$ contributes to the integral, which takes the form 
\ie 
-\hbar \int_{-\infty}^{\infty} dt' \int_0^{\infty} r_{w''} dr_{w''}  d\theta_{w''}  \left( \frac{3}{4}\frac{1}{2\pi}\right) \bigg(\frac{2
(\bar{z}'-\bar{z}'')}{((t')^2+r_{w''}^2+|z'-z''|^2)^{5/2}}\bigg) \bigg(\sum_{n\in \{0\} \cup \mathbb{N}}  J_{an} (w')^n\bigg)T_{R}^a  .
\fe 
Performing the $r_{w''}$ and $\theta_{w''}$ integrals, we arrive at
\ie 
& - 2\pi \hbar \int_{-\infty}^{\infty} dt' \left( \frac{3}{4}\frac{1}{2\pi}\right) \frac{1}{3}\bigg(\frac{2
(\bar{z}'-\bar{z}'')}{((t')^2+|z'-z''|^2)^{3/2}}\bigg)  J_{a}(w') T_{R}^a\\=& -\frac{\hbar}{z'-z''} J_{a}(w')T_{R}^a .
\fe

We thus find an expression similar to that arising at order $\hbar$ from intersecting line defects in 4d Chern-Simons theory on $\Sigma \times \mathbb{C}$, i.e., which corresponds to a rational classical R-matrix.
However, here, we note that there is dependence in the final expression on the location of the line operator, $w'$, and thus the coefficient of $\hbar/(z'-z'')$ may be regarded as a local operator. This property  reflects the fact there is topological invariance in the $t$-direction, but not on the $w$-plane. 

As a check of this result, one can also perform the computation in the holomorphic gauge $A_{\bar{z}}=0$, and arrive at the same result.
The holomorphic gauge has been used previously in the context of 4d Chern--Simons theory in \cite{Costello:2019tri}. The propagator for 5d CS in holomorphic gauge is
\ie 
\langle     A^a_t(t',z',w')  A^b_{\wb}(t'' ,z'' ,w'' )\rangle =\hbar \frac{\delta^{ab}}{z'-z''} \delta^{(2)}(w'-w'')\delta(t'-t'').
\fe 
It is then straightforward to show that 
\ie \label{fund2}
&\langle\int_{-\infty}^{\infty} dt' \int d^2w''     A^b_{\wb}(t'' ,z'' ,w'' ) J_b(w'') A^a_t(t',z',w') T_{aR}\rangle \\=  & -\hbar\int_{-\infty}^{\infty} dt' \int d^2w''      J^a(w'') T_{aR} \frac{1}{z'-z''} \delta^{(2)}(w'-w'')\delta(t'-t'')
\\=&-\frac{\hbar}{z'-z''} J_{a}(w')T_{R}^a ,
 \fe
in agreement with the result derived using the Lorentz gauge.
\subsection{Miura Operator from M2-M5 Intersection}
\label{firstmiura}

At leading order, we have derived the expression 
\ie \label{ldord}
\mathds{1} -\frac{\hbar}{z'-z''} J_{a}(w')T_{R}^a + O(\hbar^2).
\fe
This quantity is expected to satisfy a Yang-Baxter equation, as we will argue in Section \ref{rmat}. 
The Yang--Baxter equation in question is of the form 
\ie 
R_{WW'}(\tilde{z}_1,\tilde{z}_2)M_{AW}(z'-\tilde{z}_1)M_{AW'}(z'-\tilde{z}_2)=M_{AW'}(z'-\tilde{z}_2)M_{AW}(z'-\tilde{z}_1)R_{WW'}(\tilde{z}_1,\tilde{z}_2),
\fe 
where $\tilde{z}_1$ and $\tilde{z}_2$ are the positions on the $z$-plane of two surface defects, while $z'$ is the position on the $z$-plane of a line operator intersecting both surface operators. The subscripts $W$, $W'$ and $A$ indicate the algebras of local operators associated with the two surface and one line operators, respectively. The R-matrix-like operator we have derived from a line-surface intersection is identified with the operator $M$ in this equation, while $R$ is an operator that should arise from the exchange of the surface defects. Here, we see that we can multiply, say, $M_{AW}$ on both sides of the equation by some function of $\frac{z'-\tilde{z}_1}{\hbar}$, where the factor of $\hbar$ is included to preserve the symmetry of the action under simultaneous scaling of $\hbar$ and $z$.
This implies that we are allowed to modify the normalization of a solution, $M_{AW}$, of this Yang-Baxter equation, by  a function of $\frac{z'-\tilde{z}_1}{\hbar}$.

Let us compare this with 4d Chern--Simons theory. 
In 4d Chern-Simons theory, the normalization of the R-matrix can be fixed, up to some sign ambiguities, using the unitarity condition on the R-matrix. However, it is not obvious what the analogue of the unitarity condition is in 5d Chern-Simons theory.
This is because the unitarity condition in 4d Chern--Simons theory follows from the physical equivalence of two configurations of a pair of Wilson lines \cite{Costello:2017dso}, which is a consequence of the diffeomorphism invariance of 4d Chern-Simons theory along a 2-manifold. Such a symmetry is not a feature of 5d Chern-Simons theory, which only has diffeomorphism invariance along one real dimension. 

As such,  we shall motivate the choice of normalization of \eqref{ldord} using string and M-theory. As noted in \cite{Gaiotto:2020dsq}, the intersection point of the M2- and M5-brane ought to support fermion zero-modes, with mass controlled by separation between the branes. For a pair of zero-modes, the expectation value of the M2-M5 intersection will be proportional to the distance between the M2- and M5-branes along the $\mathbb{C}$-plane transverse to both branes. 
Picking the location of the surface defect, $z''$ to be zero, this distance is just proportional to $z'$ in \eqref{ldord}. 
Hence, fixing the normalization of  \eqref{ldord}  by multiplying it by $-z'/\hbar$, we find
 \ie \label{ldord2}
-\mathds{1} \frac{z'}{\hbar} +  T_{R}^a  J_{a}(w') + O(\hbar),
\fe
in agreement with  \cite{Gaiotto:2020dsq}.

Now, if we demand non-commutativity between the coordinates $z'$ and $w'$, we find that $z'$ ought to be identified with $\epsilon_2 \partial_w'$, since $[\epsilon_2\partial_{w'},w']=\epsilon_2$.\footnote{This quantization of $(z',w')$ into a Weyl algebra (due to the underlying non-commutativity of $\mathbb{C}^2$) was also used to argue that the M2-M5 intersection could be identified with the Miura operator for $G=GL(1)$ in \cite{Gaiotto:2020dsq}.} This leads us to the matrix-valued differential operator
 \ie \label{ldord3}
-\frac{\epsilon_2}{\epsilon_1}\mathds{1}\partial_{w'} +  T_{R}^a J_{a}(w') + O(\hbar),
\fe
where we have identified $\hbar$ with $\epsilon_1$.
Let us pick $R$ to be the fundamental representation of $\mathfrak{gl}_K$, whereby \eqref{ldord3} can be written as 
\ie \label{nonabelianmiura}
-\alpha\mathds{1}\partial_{w'} + E^A_{\textrm{ }B}  J^B_{\textrm{ }A}(w') + O(\hbar),
\fe 
where $E^A_{\textrm{ }B}$ is the elementary basis for $\mathfrak{gl}_K$, satisfying $
{E}_A{ }^B {E}_C{ }^D=\delta^B{ }_C {E}_A{ }^D
$, and where $\alpha =\epsilon_2/\epsilon_1$.
The first two terms are precisely of the form of an elementary Miura operator for a matrix-extended $W_{\infty}$ algebra, as defined in \cite{Eberhardt:2019xmf,Gaiotto:2023ynn}. For the case of $\mathfrak{gl}_1$, this reduces to the familiar elementary Miura operator for $W_{\infty}$ algebras 
of the form 
\ie \label{abelianmiura}
-\alpha \partial_{w'} +  J(w') + O(\hbar).
\fe 
Note that since the coupling to the Beltrami differential that describes the backreaction of the M5-branes is quadratic in the gauge fields, this coupling could generate corrections at order $\hbar$ and higher in \eqref{nonabelianmiura}, in principle.
In addition, when turning on non-commutativity, there could also be corrections to the formula \eqref{nonabelianmiura} at higher orders in the non-commutativity parameter $\epsilon_2$.

We would like to address the possible corrections at order $\hbar$ or higher order in \eqref{nonabelianmiura} (which would correspond to corrections at order $\hbar^2$ or higher prior to rescaling by $\epsilon_2/\hbar$), as well as possible higher order corrections in $\epsilon_2$, which are not part of the usual definition of the elementary Miura operator. Firstly, recall that the rational R-matrix for $\mathfrak{gl}_K$ can be written as 
\ie \label{4drmatt}
z \mathds{1}\otimes \mathds{1} + \hbar E^A_{\textrm{ }B}\otimes E^B_{\textrm{ }A}.
\fe
This quantity is derivable from intersecting Wilson lines in $\mathfrak{gl}_K$  4d Chern-Simons theory, and no higher order corrections in $\hbar$ are expected. To be precise, although there are Feynman diagrams that are higher order in $\hbar$, their  contributions can be absorbed into an overall rescaling of the R-matrix. 
The reason for this is that the color factors for the Feynman diagrams that are higher order in $\hbar$ can all be shown to be proportional to either to the identity $\mathds{1}$ or the permutation operator $E^A_{\textrm{ }B}\otimes E^B_{\textrm{ }A}$. 
Invariance of the action under a common rescaling of $\hbar$ and $z$ moreover implied that these quantities only enter via the ratio $\hbar/z$, and therefore the contributions of all Feynman diagrams reduce to an expression proportional to \eqref{4drmatt}.
For general gauge groups, it is known that the Yang-Baxter equation constrains its solutions such that the leading order contribution determines higher order contributions to all orders. We expect that the identification of the Miura operator as an R-matrix should similarly constrain contributions that are higher order in $\hbar$ and $\epsilon_2$, and this identification is explained in Section \ref{rmat}. 

We emphasize that we have provided an explicit Feynman diagram computation of the Miura operator. Although the Miura operator has been argued to arise from M2-M5 intersections before in \cite{Gaiotto:2020dsq}, an explicit Feynman diagram computation was not performed there.

\subsection{$W$-algebras from Coincident M5-branes}\label{walg}

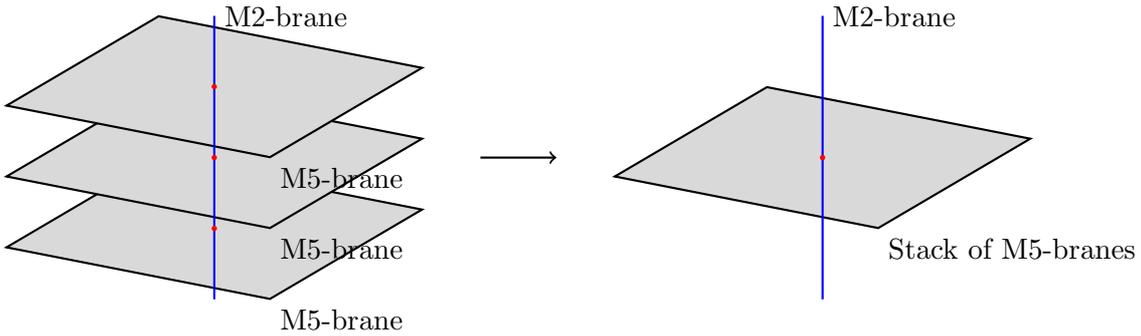
\begin{figure}
\begin{tikzpicture}

  \begin{scope}
    \tdplotsetmaincoords{70}{120}

    \begin{scope}[tdplot_main_coords]
      \fill[gray!30] (-2,-2,-1) -- (2,-2,-1) -- (2,2,-1) -- (-2,2,-1) -- cycle;
      \draw[thick] (-2,-2,-1) -- (2,-2,-1) -- (2,2,-1) -- (-2,2,-1) -- cycle;

      \fill[gray!30] (-2,-2,0) -- (2,-2,0) -- (2,2,0) -- (-2,2,0) -- cycle;
      \draw[thick] (-2,-2,0) -- (2,-2,0) -- (2,2,0) -- (-2,2,0) -- cycle;

      \fill[gray!30] (-2,-2,1) -- (2,-2,1) -- (2,2,1) -- (-2,2,1) -- cycle;
      \draw[thick] (-2,-2,1) -- (2,-2,1) -- (2,2,1) -- (-2,2,1) -- cycle;

      \draw[thick, blue] (0,0,-2) -- (0,0,2);

      \fill[red] (0,0,-1) circle (1pt);
      \fill[red] (0,0,0) circle (1pt);
      \fill[red] (0,0,1) circle (1pt);
  
      \node[below right] at (2,2,-1) {M5-brane};
      \node[below right] at (2,2,0) {M5-brane};
      \node[below right] at (2,2,1) {M5-brane};
      \node[right] at (0,0,2) {M2-brane};
    \end{scope}
  \end{scope}

  \draw[thick, ->] (3.5,0) -- +(1,0);

  \begin{scope}[shift={(8,0)}]
    \tdplotsetmaincoords{70}{120}

    \begin{scope}[tdplot_main_coords]
      \fill[gray!30] (-2,-2,0) -- (2,-2,0) -- (2,2,0) -- (-2,2,0) -- cycle;
      \draw[thick] (-2,-2,0) -- (2,-2,0) -- (2,2,0) -- (-2,2,0) -- cycle;

      \draw[thick, blue] (0,0,-2) -- (0,0,2);

      \fill[red] (0,0,0) circle (1pt);
  
      \node[below right] at (2,2,0) {Stack of M5-branes};
      \node[right] at (0,0,2) {M2-brane};
    \end{scope}
  \end{scope}

\end{tikzpicture}
\caption{Miura transformation from stacking M5-branes }
\label{miurastack}
\end{figure}

Given an elementary Miura operator for a (matrix-extended) $W_{\infty}$-algebra, one can concatenate several copies of it to produce a Miura transformation, which is a standard method for constructing $W$-algebras. This transformation is given explicitly by 
\begin{equation}\label{mtrans}
\mathcal{L}(z)=\prod_{i=1}^n\left(-\alpha \mathds{1} \partial_w+J^{(i) A}{ }_B(w) {E}_A{ }^B\right)=\sum_{k=0}^n U_{(k)}{ }^A{ }_B(w) {E}_A{ }^B(-\alpha \partial_w)^{n-k},
\end{equation}
where $U_{(0)}{ }^A{ }_B(w)=1$, as described in \cite{Eberhardt:2019xmf}. The coefficients of the differential operator on the RHS generate the (matrix-extended) $\mathcal{W}_n$ algebras, and the (matrix-extended) ${W}_{\infty}$ algebra arises from taking $n\rightarrow \infty $. From the perspective of twisted M-theory, this corresponds to stacking M5-branes that intersect with an M2-brane as depicted in Figure \ref{miurastack}. In fact, this stacking equips the (matrix-extended) $\mathcal{W}_{\infty}$ algebra with a coproduct structure, and we shall derive this coproduct in  Section \ref{m5stack}.

For the case of $K=1$, this observation was first made in \cite{Gaiotto:2020dsq}, and is consistent with the AGT correspondence \cite{Alday:2009aq}. However, we would like to understand the general Miura transformation explicitly in terms of a Feynman diagram computation.

Here, we point out that the construction of the general Miura operator is analogous to how monodromy matrices are constructed in 4d Chern-Simons theory in \cite{Costello:2018gyb}. 
Recall that for 4d CS, a general Wilson line that is classically in a representation of $\mathfrak{g}[[z]]$, where $\mathfrak{g}=\mathfrak{gl}_K$, has $\frac{1}{k!} \partial_z^k A$ coupled to an element $t^A{ }_B[k]$ of $\mathfrak{g}[[z]]$, where $k=0,\ldots, n$ for some positive integer $n$. As explained in \cite{Costello:2018gyb}, when crossing such a Wilson line with an ordinary Wilson line (in a representation of $\mathfrak{g}$), the resulting R-matrix at leading order in $\hbar$ takes the form
\ie \label{cw2}
r= \sum_{k \geq 0}^n \frac{1}{k!} \partial_z^k \frac{1}{(-z)}\left(t^A{ }_B[k] \otimes E_A{ }^B\right)=- \sum_{k \geq 0}^n(-1)^k \frac{1}{z^{k+1}}\left(t^A{ }_B[k] \otimes E_A{ }^B\right).
\fe

Now, observe that by multiplying \eqref{cw2} by $(-z)^{n+1}$, we obtain
\ie 
\sum^n_{k\geq 0} ( t^A{ }_B[k] \otimes E_A{ }^B (-z)^{n-k}).
\fe 
This is analogous in form to the RHS of \eqref{mtrans}, which can be seen by replacing $t^A{ }_B[k]$ by the local operator $U_{(k)}{ }^A{ }_B(w) $, and $z$ by $\alpha \partial_w$. If we generalize the computation given in  Section \ref{2.2} (with $\mathfrak{g}=\mathfrak{gl}_K$) such that the surface operator is now of the form \eqref{gensurf}, i.e., by employing a surface operator with couplings of $\frac{\hbar^k}{k!} \partial_z^k A_{\bar{w}}$ to higher spin currents $U_{(k)}{ }^A{ }_B(w) $ (where $k=0,\ldots, n$, and where the factor of $\hbar^k$ is necessary to preserve the symmetry of the action under simultaneous rescaling of $\hbar$ and $z$), while taking non-commutativity into account, we would obtain the Miura transformation given on the RHS of \eqref{mtrans}. Indeed, performing the explicit computation with the general surface defect at $z=0$, we find 
\ie \label{fundgen}
&\langle\int_{-\infty}^{\infty} dt' \int d^2w''    \frac{\hbar^k}{k!} \partial^k_{z'} [(A_{\wb})_ A^{\textrm{  }B}(t'' ,0 ,w'' )] U_{(k)}{ }^A{ }_B(w'') [ (A_t)^C_{\textrm{  }D}(t',z',w') ]E_C^{\textrm{  }D}\rangle\\=& -\hbar \int_{-\infty}^{\infty} dt' \int d^2w''   \left( \frac{3}{4}\frac{1}{2\pi}\right) \frac{\hbar^k}{k!} \partial^k_{z'}\bigg(\frac{2
(\bar{z}')}{((t'-t'')^2+|w'-w''|^2+|z'|^2)^{5/2}}\bigg)U_{(k)}{ }^A{ }_B(w''){E}_A{ }^B 
\\=& -\hbar^{k+1} (-1)^k \frac{1}{(z')^{k+1}}
U_{(k)}{ }^A{ }_B(w''){E}_A{ }^B .
\fe
Summing over $k=0,\ldots ,n$ with a normalization of $(-\frac{z'}{\hbar})^{n+1}$, we arrive at 
\ie 
\sum_{k=0}^n U_{(k)}{ }^A{ }_B(w'') {E}_A{ }^B\left(-\frac{z'}{\hbar}\right)^{n-k}
\fe 
and identifying $\frac{z'}{\hbar}$ with $\alpha\partial'_w$ as we did in Section 
\ref{firstmiura},  we arrive at the RHS of \eqref{mtrans}. The 
 surface operators coupled to holomorphic derivatives of the gauge field indeed arise naturally from stacking M5-branes, as we shall derive in Section \ref{m5stack}.

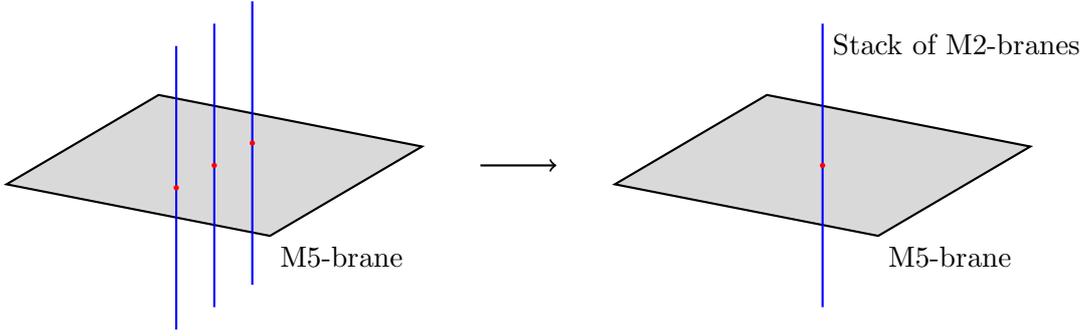
\begin{figure}
\begin{tikzpicture}

  \begin{scope}
    \tdplotsetmaincoords{70}{120}

    \begin{scope}[tdplot_main_coords]
      \fill[gray!30] (-2,-2,0) -- (2,-2,0) -- (2,2,0) -- (-2,2,0) -- cycle;
      \draw[thick] (-2,-2,0) -- (2,-2,0) -- (2,2,0) -- (-2,2,0) -- cycle;

      \draw[thick, blue] (-1,0,-2) -- (-1,0,2);

      \draw[thick, blue] (0,0,-2) -- (0,0,2);

      \draw[thick, blue] (1,0,-2) -- (1,0,2);
  
      \fill[red] (-1,0,0) circle (1pt);
      \fill[red] (0,0,0) circle (1pt);
      \fill[red] (1,0,0) circle (1pt);

      \node[below right] at (2,2,0) {M5-brane};
    \end{scope}
  \end{scope}

  \draw[thick, ->] (3.5,0) -- +(1,0);

  \begin{scope}[shift={(8,0)}]
    \tdplotsetmaincoords{70}{120}

    \begin{scope}[tdplot_main_coords]
      \fill[gray!30] (-2,-2,0) -- (2,-2,0) -- (2,2,0) -- (-2,2,0) -- cycle;
      \draw[thick] (-2,-2,0) -- (2,-2,0) -- (2,2,0) -- (-2,2,0) -- cycle;

      \draw[thick, blue] (0,0,-2) -- (0,0,2);

      \fill[red] (0,0,0) circle (1pt);
  
      \node[below right] at (2,2,0) {M5-brane};
      \node[below right] at (0,0,2) {Stack of M2-branes};
    \end{scope}
  \end{scope}

\end{tikzpicture}
\caption{Stacking M2-branes}
\label{miurastack2}
\end{figure}

As explained in \cite{Gaiotto:2020dsq}, one can also take the OPEs of Wilson lines piercing a surface operator at different points on the $w$-plane, and form a single Wilson line, as depicted in \ref{miurastack2}. For $G=GL(1)$, this operation furnishes Calogero-like differential operators in the variables $w_i$, which denote the original intersection points of the M2-branes. 
\section{Coproducts from M2- and M5-branes}\label{sec3}

In 4d Chern--Simons theory, the OPE of Wilson lines furnishes the coproduct of the Yangian \cite{Costello:2017dso}, and is one way to deduce that line operators in 4d Chern--Simons theory are associated with representations of the Yangian algebra. This fact may also be deduced from computing quantum corrections to gauge invariance of Wilson lines as shown in \cite{Costello:2017dso}, or deriving RTT-type relations satisfied by the line operators as in \cite{Costello:2018gyb}. 
 It is thus natural to expect that the coproduct
associated with operators in 5d Chern--Simons theory ought to arise from OPEs of these operators. Computations of such OPEs for abelian 5d Chern--Simons theory were in fact performed in \cite{Oh:2021wes}, producing the coproduct associated with the affine Yangian of $GL(1)$.

In nonabelian 5d Chern--Simons theory, one-loop quantum corrections to the gauge invariance of a general Wilson line  associates it with a representation of the deformed double current algebra \cite{Costello:2017fbo}, denoted $\mathcal{A}^{(K)}$, while general surface defects are associated with a matrix-extended $W_{\infty}$ algebra, denoted $\mathcal{W}^{(K)}_{\infty}$.
In this section, we shall derive  coproducts involving these algebras from stacking of M2-branes, stacking of M5-branes, and demanding the gauge invariance of M2-M5 intersections. We shall show that the OPEs of operators associated with M2-branes and M5-branes are non-singular and well-defined, leading to well-defined fusion of said operators. This underlies the Yang-Baxter equation that has the Miura operator as its solution.

As we shall see, we shall find matches with coproducts derived using mathematically rigorous methods by Gaiotto, Rapcak and Zhou in \cite{Gaiotto:2023ynn}.
We shall turn off non-commutativity first, and focus  on coproducts in the commutative case. Coproducts in the case of $GL(1)$ non-commutative 5d Chern-Simons theory were already computed in \cite{Oh:2021wes}, so we shall not delve into details of computations for the non-commutative case, and instead focus on computations for the $SL(K) \subset GL(K)$ subgroup. However, we shall  explain how the $GL(1)$ results are generalized by a new interaction term present when the gauge group is replaced by $GL(K)$, which mixes $GL(1)$ and $SL(K)$ gauge fields.

In what follows we shall only work to leading order in $\hbar$. The one-loop exactness of these results is expected to follow from arguments similar to those given in \cite{Oh:2021wes}.  In Section 3.8 of this work, the 1-loop exactness of Feynman diagrams used to compute coproducts in abelian 5d Chern--Simons theory was shown. In Figures 12 and 13 of \cite{Oh:2021wes}, possible higher-loop diagrams that could modify diagrams of the sort that we will consider below were considered. We should consider analogous diagrams in the present case, and show that they do not contribute. These diagrams involve cyclic wedge products of propagators.  These cyclic wedge products can be shown to vanish, that is, defining $P_{ij}:=P(t_i-t_j,w_i-w_j,z_i-z_j)$,  it can be shown that 
\ie 
P_{12}\wedge P_{21} =0
\fe 
and 
\ie 
P_{12}\wedge P_{23}\wedge P_{31} =0.
\fe
These identities can be proven via explicit computation, or by a straightforward generalization of the proof for nonabelian 4d Chern--Simons theory in Appendix C.1 of \cite{Ishtiaque:2018str}. 
  Hence, higher-loop effects do not modify our analyses below.
   \subsection{Fusion of M2-branes }

   We would like to first compute the OPE of Wilson lines in 5d nonabelian Chern-Simons theory, in analogy to the computation of \cite{Costello:2017dso} for 4d Chern-Simons theory.

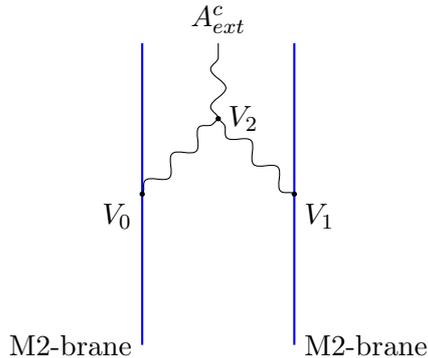
\begin{figure}
\begin{center}
\begin{tikzpicture}

  \draw[thick, blue] (-1,2) -- (-1,-2);
  \fill[black] (-1,0) circle (1pt);
  \node[below left] at (-1,0) {\(V_0\)};

  \draw[thick, blue] (1,2) -- (1,-2);
  \fill[black] (1,0) circle (1pt);
  \node[below right] at (1,0) {\(V_1\)};

  \draw[decorate, decoration={snake, amplitude=1mm, segment length=5mm}] (-1,0) -- (0,1);
  \draw[decorate, decoration={snake, amplitude=1mm, segment length=5mm}] (1,0) -- (0,1);

  \draw[decorate, decoration={snake, amplitude=1mm, segment length=5mm}] (0,1) -- (0,2);
  \node[above] at (0,2) {\(A^c_{{ext}}\)};

  \fill[black] (0,1) circle (1pt);
  \node[below right] at (0,1.3) {\(V_2\)};

  \node[right] at (1,-2) {M2-brane};
  \node[left] at (-1,-2) {M2-brane};

\end{tikzpicture}
\end{center}
\caption{M2-brane fusion in nonabelian 5d Chern-Simons theory.}
\label{fig1222e}
\end{figure}

Here, we shall compute the OPE of Wilson lines separated along the complex $w$-plane by a distance $\tilde{w}$, as depicted in Figure \ref{fig1222e}. 
   The bulk interaction vertex for nonabelian 5d Chern-Simons theory with non-commutativity turned off is 
   \ie 
-\frac{1}{4\pi}f^{abc}dw\wedge dz,
   \fe
   where $a$, $b$, and $c$ denote $SL(K)$ indices. 
To compute the OPE of ordinary Wilson lines, we would like to compute the integral 
\ie \label{integg}
-\frac{1}{4\pi}t^a\otimes t^bf_{abc} \int_{t_1} \int_{t_2}\int_{t,w,\bar{w},z,\bar{z}}P(t_1-t,w,z) \wedge dw\wedge dz \wedge A_{ext}^c(t,w,z) \wedge P(t_2-t, w-\tilde{w},z),
\fe 
where $A^c_{ext}$ denotes an external gauge field. The computation of this integral, which uses the techniques of \cite{Costello:2017dso}, can be found in Appendix \ref{A}, with the result that we obtain a Wilson line of the form 
\label{integdline}
\ie 
 &\frac{\hbar }{\pi \tilde{w}} t^a\otimes t^b f_{abc} \int dt \partial_z A_{ext}^c(t,w,\bar{w},z,\bar{z}).
\fe

We could have as well considered two Wilson lines separated along the $z$-plane instead of the $w$-plane. The fusion of two such Wilson lines would give a Wilson line coupled to $\partial_w A$ instead of $\partial_z A$.  
Thus, in general, a Wilson line has a coupling of the $\partial^m_z\partial^n_w A$ for some non-negative integers $m$ and $n$, and is considered to be in a representation of the double polynomial loop algebra $\mathfrak{g}[[w,z]]$ at the classical level, as given in \eqref{wilson}.

We would like to generalize the computation described above to the case of fusing Wilson lines with couplings $\frac{1}{m!}\partial^m_w A$ and $\frac{1}{n!}\partial^n_w A$ for some positive integers $m$ and $n$. This shall lead us to the  meromorphic coproduct
\ie 
\Delta_{\mathcal{A}} (\tilde{w}):\mathcal{A}^{(K)}\rightarrow \mathcal{A}^{(K)} \otimes \mathcal{A}^{(K)} ((\tilde{w}^{-1}))
\fe
for the deformed double current algebra $\mathcal{A}^{(K)}$. The main coproduct of interest that we shall aim to reproduce takes the form 
\begin{equation}\label{coprod1}
\begin{aligned}
& \Delta_{\mathcal{A}}(\tilde{w})\left(t^a_{1,0}\right)=
t^a_{1,0}\otimes 1+1 \otimes t^a_{1,0}
 + C_1\epsilon_1 \sum_{m, n \geq 0} \frac{(-1)^{n}}{\tilde{w}^{m+n+1}}\frac{(m+n)!}{m!n!} f^{a }_{\textrm{ }bc} t^b_{0,m}t^c_{0,n} 
\end{aligned}
\end{equation}
for some constant $C_1$.

The explicit Feynman diagram computation is carried out in Appendix \ref{genline}.
The result is that the line operator resulting from fusion  has a  coupling to the external gauge field of the form
\ie
 \hbar f_{abc}\sum_{m, n \geq 0}{c_{m,n}}t^a_{0,m}t^b_{0,n}\tilde w^{-m-n-1}\int_{\bR_t}\pa_{z_2}A^c_{ext},
\fe
where
\ie
c_{m,n}=\frac{1}{\pi }(-1)^{n} \frac{(m+n)!}{m!n!}.
\fe
This result agrees with the coproduct \eqref{coprod1} derived in \cite{Gaiotto:2023ynn}. 

There is another coproduct involving fusion of Wilson lines to a Wilson line coupled to the $GL(1)$ gauge field. Denoting the modes coupled to the $GL(1)$ gauge field as $t_{m,n}$, the coproduct takes the form 
\begin{equation}\label{coprod1b}
\begin{aligned}
\Delta_{\mathcal{A}}(\tilde{w})\left(t_{2,0}\right)=&
t_{2,0}\otimes 1+1 \otimes t_{2,0}\\&
 + C_2\epsilon_1 \epsilon_2 \sum_{m, n \geq 0} \frac{(-1)^{n}}{\tilde{w}^{m+n+2}}\frac{(m+n+1)!}{m!n!} \left(t_{a,0,m}\otimes t^a_{0,n}  + \frac{\epsilon_1}{\epsilon_2}t_{0,m}\otimes t_{0,n} \right),
\end{aligned}
\end{equation}
for some constant $C_2$. This coproduct can be derived by a straightforward generalization of the derivation in \cite{Oh:2021wes}, taking into account the
interaction term mixing the $GL(1)$ and $SL(K)$ gauge fields which has the form 
\ie \label{mixint}
A\wedge \partial_z A^{a}\wedge \partial_w A_{a},
\fe
where the gauge field with no index corresponds to the $GL(1)$ subgroup of the gauge group.

\subsection{Fusion of M5-branes}\label{m5stack}

\begin{figure}
\begin{center}
\begin{tikzpicture}[tdplot_main_coords]

  \fill[gray!30] (-2,-2,-1) -- (2,-2,-1) -- (2,2,-1) -- (-2,2,-1) -- cycle;
  \draw[thick] (-2,-2,-1) -- (2,-2,-1) -- (2,2,-1) -- (-2,2,-1) -- cycle;

  \fill[gray!30] (-2,-2,1) -- (2,-2,1) -- (2,2,1) -- (-2,2,1) -- cycle;
  \draw[thick] (-2,-2,1) -- (2,-2,1) -- (2,2,1) -- (-2,2,1) -- cycle;

  \draw[decorate, decoration={snake, amplitude=1mm, segment length=5mm}] (0,-1,-1) -- (1,1,0);
  \draw[decorate, decoration={snake, amplitude=1mm, segment length=5mm}] (0,-1,1) -- (1,1,0);

  \draw[decorate, decoration={snake, amplitude=1mm, segment length=5mm}] (1,1,0) -- (1,3,0);
  \node[right] at (2.1,2.3,0) {\(A^c_{{ext}}\)};

  \fill[black] (1,1,0) circle (1pt);

  \fill[black] (0,-1,-1) circle (1pt);
  \node[above left=2pt] at (1.4,0,-1) {\(V_1\)};
\node[above left=2pt] at (1,-1.7,-1) {\(t=0\)};

  \fill[black] (0,-1,1) circle (1pt);
  \node[above left=2pt] at (1.4,0,1) {\(V_2\)};
\node[above left=2pt] at (1,-1.7,1) {\(t=\epsilon\)};
\end{tikzpicture}
\end{center}
\caption{M5-brane fusion in nonabelian 5d Chern-Simons theory.}
\label{fig_rotated}
\end{figure}
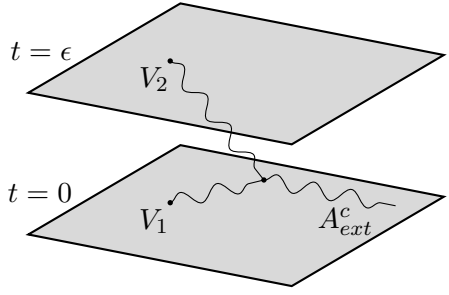

We would like to now compute the OPE of surface operators in nonabelian 5d Chern-Simons theory, as depicted in Figure \ref{fig_rotated}.
Notably this shall define the coproduct of the matrix-extended $\mathcal{W}_{\infty}$-algebra, which maps $\mathcal{W}_{\infty}^{(K)}\rightarrow \mathcal{W}_{\infty}^{(K)} \otimes \mathcal{W}_{\infty}^{(K)}$.

The integral we ought to compute takes the form\footnote{The prefactor of $\frac{1}{(-i2)^2}$ is necessary since we define the measure on the surface operator to be $\frac{dw d\bar{w}}{(-i2)}$.} 
\ie \label{surffuseampl}
&-\frac{1}{(-i2)^2}\int_{w_1}  \int_{w_2} \int_{t,w,\bar{w},z,\bar{z}}\frac{\hbar}{4\pi}  J^a(w_1)\otimes \tilde{J}^b(w_2) f_{abc}\\&\times   P(t,w_1-w,z)\wedge dw\wedge dz \wedge A_{ext}^c(t,w,z)\wedge P(t-\epsilon,w_2-w,z),
\fe 
where we have used the notation $\tilde{J}$ to distinguish between the currents of the two surface operators.
The explicit computation is carried out in Appendix \ref{surffuse}, and the result is 
\ie 
\frac{\hbar}{4\pi }\int_{w,\bar{w}} J^a(w)\otimes \tilde{J}^b(w) f_{abc} dw \wedge \partial_z A_{ext}^c(t,w,z) .
\fe 
This surface operator arising from fusion notably involves a holomorphic derivative of the gauge field, and a spin-2 current. 

More generally, one can show that two surface operators coupled to $\partial^m_zA_{\bar{w}} $ (and a current of spin $m+1$)
and $\partial^n_zA_{\bar{w}} $ (and a current of spin $n+1$) respectively fuse to a surface operator coupled to $\partial^{m+n+1}_zA_{\bar{w}} $, and to a current of spin $m+n+2$. This coproduct structure is of the expected form, and agrees with the coproduct structure of $\mathcal{W}^{(K)}_{\infty}$ described in \cite{Gaiotto:2023ynn}. The similarity between the computation given here and the computation of the Yangian coproduct in \cite{Costello:2017dso}
can be attributed to the fact that there exists an algebra isomorphism between the Zhu algebra of $\mathcal{W}^{(K)}_{\infty}$ and the Yangian of $\mathfrak{gl}_K$ \cite{Gaiotto:2023ynn}. The computation above can be generalized to take into account non-commutativity, by generalizing the computation of \cite{Oh:2021wes} while taking into account the mixed interaction term \eqref{mixint}, and the conclusion that surface operators supporting currents of fixed spin can be fused to produce surface operators supporting higher spin currents is unchanged.

\subsection{M2-M5 Coproduct}
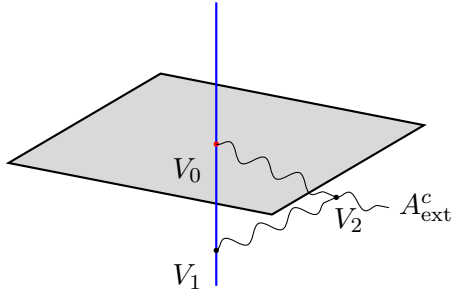
\begin{figure}
\begin{center}
\begin{tikzpicture}[tdplot_main_coords]

  \fill[gray!30] (-2,-2,0) -- (2,-2,0) -- (2,2,0) -- (-2,2,0) -- cycle;
  \draw[thick] (-2,-2,0) -- (2,-2,0) -- (2,2,0) -- (-2,2,0) -- cycle;

  \draw[thick, blue] (0,0,-2) -- (0,0,2);

  \fill[red] (0,0,0) circle (1pt);
  \fill[black] (0,0,-1.5) circle (1pt);
  \fill[black] (1,2.4,0) circle (1pt);
  
  \node[below left=2pt] at (0,0,0) {\(V_0\)};
  \node[below left=2pt] at (0,0,-1.5) {\(V_1\)};
  \node[below right=2pt] at (1,2.15,0) {\(V_2\)};

  \draw[decorate, decoration={snake, amplitude=1mm, segment length=5mm}] (0,0,-1.5) -- (1,2.4,0);
  \draw[decorate, decoration={snake, amplitude=1mm, segment length=5mm}] (0,0,0) -- (1,2.4,0);

  \draw[decorate, decoration={snake, amplitude=1mm, segment length=5mm}] (1,2.4,0) -- (1,3.2,0);
  \node[right] at (1,3.2,0) {\(A^c_{\text{ext}}\)};

\end{tikzpicture}
\end{center}
\caption{Mixed coproduct from M2-M5-brane intersection.}
\label{figmixed}
\end{figure}

We shall now compute the mixed coproduct that involves both $\mathcal{A}^{(K)}$ and $\mathcal{W}_{\infty}^{(K)}$. Unlike the previous coproducts, this does not have a simple description in terms of fusion of line or surface defects, but rather arises from a gauge-invariance constraint at the junction of a Wilson line and surface operator, as explained in \cite{Gaiotto:2020dsq} and depicted in Figure \ref{figmixed}. For the case of $GL(1)$, the gauge-invariance constraint relates the modes of the $W$-algebra currents and the local operators on the Wilson line as 
\begin{equation}
t_{n, m}^{\mathrm{up}} \cdot O+W_{m-n}^{(n+1)} \cdot O+(\ldots) \cdot O-O \cdot t_{n, m}^{\text {down }}=0,
\end{equation}
where $O$ denotes the junction between the surface and line operators, $W^{(n+1)}_{ m-n} \equiv \oint z^m W_{n+1}(z)$ is a mode of the spin $n+1$ current $W^{(n+1)}$, where the ellipsis indicates possible quantum corrections, and where ``up" and ``down" denote local operators of the Wilson line above and below the surface operator, respectively. We expect the mixed coproduct to similarly arise from an M2-M5 intersection for the case of nonabelian gauge group. 

\tdplotsetmaincoords{70}{120}

The mixed coproduct in the present context is the map  $\mathcal{A}^{(K)}\rightarrow \mathcal{A}^{(K)} \otimes \mathcal{W}_{\infty}^{(K)}$, denoted as $\Delta_{\infty}$ in \cite{Gaiotto:2023ynn}. Our primary focus will be the coproduct that takes the form
\begin{equation}\label{exmixcop}
\Delta_{\infty} (t^a_{1,0})= t^a_{1,0}\otimes 1 + 1 \otimes \Psi_{\infty}(t^a_{1,0}) +C_3\epsilon_1 f^a_{\textrm{ }bc}\sum_{n=0} t^b_{0,n}\otimes J^c_{-n-1}, \end{equation}
where $C_3$ is a numerical constant, $J^c_{-n-1}$ is a mode of a spin-1 current, and $\Psi_{\infty}$ is a map from $\mathcal{A}^{(K)}$ to $\mathcal{W}^{(K)}_{\infty}$. We emphasize that $C_3$ is independent of $n$, and therefore the result of the evaluation of the relevant Feynman amplitude should have no $n$-dependence either.

The Feynman amplitude is evaluated explicitly in Appendix \ref{linesurf}, where it is shown that the numerical coefficient has no dependence on $n$, that is we find
\begin{equation}
\sum_n c  f_{abc}J^a_{-n-1} t^b_{0, 1} \partial_{z_2} A^c
\end{equation}
where 
\ie 
c=-\frac{\hbar}{8\pi^2}.
\fe
This agrees with the mixed coproduct for $\mathcal{A}^{(K)}$ and $\mathcal{W}_{\infty}^{(K)}$ given in \eqref{exmixcop}.

We can also compute the mixed coproduct which results in a Wilson line coupled to the $GL(1)$ gauge field, which takes the form
\begin{equation}\label{abnomix}
\Delta_{\infty}\left({t}_{2,0}\right)={t}_{2,0}\otimes 1 + 1\otimes \Psi_{\infty}({t}_{2,0})  + C_4  \epsilon_1 \epsilon_2 \sum_{n=1}^{\infty} n\left(t^a_{0, n-1} \otimes J_{a,-n-1}+ \frac{\epsilon_1}{\epsilon_2}{t}_{0, n-1} \otimes J_{-n-1}\right),
\end{equation}
where $C_4$ is a numerical constant. This can be achieved by generalizing the computation of \cite{Oh:2021wes} while taking into account the mixed interaction term \eqref{mixint}.  The crucial difference is that this interaction 
term contributes a cubic interaction vertex relating a $GL(1)$ gauge field to two $SL(K)$ gauge fields with indices $a$ and $b$, of the form 
\ie 
\epsilon_2 \delta^{ab}\partial_z\partial_w .
\fe 
 There is thus  a new type of Feynman diagram containing such an interaction that ought to be taken into account, which gives rise to an amplitude that generalizes that of equation (69) of \cite{Oh:2021wes}, that is 
\ie 
\begin{aligned}
 t^b_{0, n-1}J^a_{-n-1}\delta_{ab} \int_{V_2} A_{\text {ext }} d z_2 d w_2 & \int_{V_0} \delta\left(t_0\right) \delta^{(2)}\left(z_0\right) \partial_{z_2}\left(w_0^n P_{02}\right) d w_0 \\
& \times \int_{V_1} \delta\left(t_1-\epsilon\right) \delta^{(2)}\left(z_1\right) \delta^{(2)}\left(w_1\right) \partial_{w_2} \partial_{w_1}^{n-1} P_{12}
\end{aligned}
\fe 
in the notation of \cite{Oh:2021wes}.
The only difference from equation (69) of \cite{Oh:2021wes} is the color factor $ t^b_{0, n-1}J^a_{-n-1}\delta_{ab} $, and the evaluation of the amplitude otherwise goes through as in \cite{Oh:2021wes}, with a result of the form
\ie 
C\epsilon_1\epsilon_2 \sum_{n=1}^{\infty} n  t^b_{0, n-1}J_{a,-n-1}, 
\fe 
for some number $C$ (having identified $\epsilon_1$ with $\hbar$ as before)
in agreement with the first term in the parenthesis in \eqref{abnomix}.
\subsection{The M2-M5 Intersection as an R-matrix }\label{rmat}

Having derived various coproducts for the algebras $\mathcal{A}^{(K)}$ and $\mathcal{W}_{\infty}^{(K)}$, we are naturally led to the question of whether these coproducts can be used to derive a Yang-Baxter equation for the Miura operators derived in Section \ref{sec2}, by relating the Miura operators to a universal R-matrix. We shall explain why this is the case in what follows.

The arguments that follow have appeared before in the recent work of Haouzi and Jeong \cite{Haouzi:2024qyo}, which studied the quantum toroidal algebra of $\mathfrak{gl}(1)$. However, we emphasize that this work did not provide a Feynman diagram derivation of the associated Miura operators from 5d Chern-Simons theory.

 Firstly, when one attempts a physical realization of Koszul duality in 5d CS (following, e.g., \cite{Paquette:2021cij}), one notes that the coupling of each defect in the theory arises from a descent procedure applied to a combination of local operators on the defect and bulk local operators (modes of the ghost
field) of 5d CS. The BRST-invariance of such couplings leads
 one to deduce that there are homomorphisms
between the Koszul dual of the algebra of bulk local operators  at the locus of line and surface defects and
the algebra of local operators on each defect. The Koszul dual of the algebra of bulk local operators will be denoted $\mathcal{A}_{\epsilon_1,\epsilon_2}$. The Miura operator, which we shall denote as $R_{\mathcal{A}\mathcal{W}}$, can be identified with  a universal R-matrix $\mathcal{R} \in \mathcal{A}_{\epsilon_1,\epsilon_2} \otimes  \mathcal{A}_{\epsilon_1,\epsilon_2}$ subject to these homomorphisms. That is, if the homomorphisms for line and surface operators are respectively 
\ie 
\rho_{\mathrm{M} 2}: \mathcal{A}_{\varepsilon_{1, \varepsilon_2}} \rightarrow {M} 2, \quad \rho_{\mathrm{M} 5}: \mathcal{A}_{\varepsilon_1, \varepsilon_2} \rightarrow {M} 5,
\fe
where $M2$ and $M5$ indicate the algebra of local operators on the line and surface defects, respectively,
then 
\ie 
R_{\mathcal{AW}}=\rho_{M2}\otimes \rho_{M5} \mathcal{R}.
\fe

In Section \ref{walg}, we have explained how concatenation of elementary Miura operators give rise to a general Miura operator.
We can identify this operation with the relation
\ie
(\mathrm{id} \otimes \Delta) (\mathcal{R})={\mathcal{R}}_{13} {\mathcal{R}}_{12},
\fe
since the concatenation of elementary Miura operators can be identified with the concatenation of operators $R_{\mathcal{AW}}$ and $R_{\mathcal{AW'}}$ associated with two different surface defects.
Also, as explained in \cite{Gaiotto:2020dsq} and Section \ref{walg}, one can also consider concatenation of multiple line operators piercing a single surface defect by taking their OPEs, which we have studied in the previous section. This would correspond to 
\ie 
(\Delta \otimes \mathrm{id}) {\mathcal{R}}={\mathcal{R}}_{13} {\mathcal{R}}_{23}.
\fe

Finally, the gauge-invariance constraint of the M2-M5 intersection provides us with a relation involving the mixed coproduct, of the form
\ie 
\mathcal{R} \Delta(g)=\Delta^{\mathrm{op}}(g) \mathcal{R},
\fe
where $\Delta^{\textrm{op}}=\sigma \cdot \Delta$ is the opposite coproduct (where $\sigma (g_1\otimes g_2) =g_2 \otimes g_1$), and $g\in \mathcal{A}_{\epsilon_1,\epsilon_2}$. 
We note that in the work of \cite{Gaiotto:2023ynn}, the Miura operator that we derived from nonabelian 5d Chern-Simons theory was shown to intertwine the coproduct and opposite coproduct in this manner.

Hence, it follows that the universal R-matrix, $\mathcal{R}$, satisfies a Yang-Baxter equation of the form 
\ie
\mathcal{R}_{12} \mathcal{R}_{13} \mathcal{R}_{23}=\mathcal{R}_{23} \mathcal{R}_{13} \mathcal{R}_{12}.
\fe
A particular instance of this Yang-Baxter equation
for the case of $GL(1)$, of the form 
\ie 
R_{\mathcal{A}\mathcal{W}}R_{\mathcal{A}\mathcal{W}'} R_{\mathcal{W}\mathcal{W}'} =R_{\mathcal{W}\mathcal{W}'} R_{\mathcal{A}\mathcal{W}'} R_{\mathcal{A}\mathcal{W}},
\fe 
is well-known to describe the exchange of ordering of Miura operators in terms of the Maulik-Okounkov R-matrix $R_{\mathcal{W}\mathcal{W}'}$ \cite{maulik2012quantum,Prochazka:2019dvu}, which in the physical picture of twisted M-theory corresponds to the exchange of two M5-branes intersecting a single M2-brane.

\appendix 

\section{Feynman Diagram Evaluation  for Fusion of Wilson Lines in Representations of $\mathfrak{g}$}\label{A}

We would like to evaluate the integral \eqref{integg}, by generalizing the computation of OPEs of Wilson lines in representations of $\mathfrak{g}$ in 4d Chern-Simons theory presented in \cite{Costello:2017dso}. In what follows, we shall replace the variable $\tilde{w}$ in \eqref{integg} by $\epsilon$. 
Noting that 
\ie 
\int^{\infty}_{-\infty}dt \frac{1}{(t^2+|w|^2+|z|^2)^{\frac{5}{2}}} = \frac{4}{3} \frac{1}{(|w|^2+|z|^2)^2},
\fe 
we can compute the integrals over $t_1$ and $t_2$ to give 
\ie \label{integd}
-\frac{1}{4\pi}t^a\otimes t^bf_{abc} \int_{t,w,\bar{w},z,\bar{z}}P'(w,z) \wedge dw\wedge dz \wedge A^c(t,w,z) \wedge P'(w-\epsilon,z),
\fe 
where we have defined the four-dimensional one-form propagator without color indices on $\mathbb{C}^2$ as 
\ie 
P^{\prime}(w, z)=-\mathrm{d} \bar{z} \partial_w Q(w, z)+ \mathrm{~d} \bar{w} \partial_z Q(w, z) 
\fe 
where 
\ie 
Q(w,z):= -\frac{1}{\pi} \frac{1}{\left(|w|^2+|z|^2\right)} .
\fe
We can rewrite \eqref{integd} as 
\ie\label{integd2}
 &-\frac{1}{4\pi } t^a\otimes t^b f_{abc} \int_{t,w,\bar{w},z,\bar{z}}A^c(t,w,z) \wedge d w \wedge dz \wedge d \bar{w} \wedge d \bar{z}\left(\partial_z Q(w) \partial_w Q(w-\epsilon)-\partial_w Q(w) \partial_z Q(w-\epsilon)\right)
\fe
where we have suppressed the $z$-dependence in $Q$. The expression \eqref{integd2} can be further rewritten as 
\ie\label{integd3}
 &-\frac{1}{4\pi } t^a\otimes t^b f_{abc} \int_{t,w,\bar{w},z,\bar{z}}A^c(t,w,z) \wedge d w \wedge dz \wedge d \bar{w} \wedge d \bar{z} \left( \partial_z (a) + \partial_w (b)\right)
\fe
where 
\ie 
(a)=&\frac{1}{2}\left(-\partial_wQ(w) Q(w-\epsilon)+Q(w)\partial_wQ(w-\epsilon)\right) = \frac{1}{2\pi^2} \frac{-\epsilon \bar{w}^2 +|\epsilon|^2\bar{w} +\bar{\epsilon} |z|^2}{(|w|^2+|z|^2)^2(|w-\epsilon|^2+|z|^2)^2}\\
(b)=&\frac{1}{2}\left(\partial_zQ(w) Q(w-\epsilon)-Q(w)\partial_zQ(w-\epsilon)\right)  =-\frac{1}{2\pi^2}\frac{\bar{z} (|w-\epsilon|^2-|w|^2)}{(|w|^2+|z|^2)^2(|w-\epsilon|^2+|z|^2)^2}.
\fe

 The integral involving $(a)$ can be performed with the aid of the identity 
\ie \label{feyn}
\frac{1}{A^\alpha B^\beta}=\frac{\Gamma(\alpha+\beta)}{\Gamma(\alpha) \Gamma(\beta)} \int_0^1 \mathrm{~d} x \frac{x^{\alpha-1}(1-x)^{\beta-1}}{(x A+(1-x) B)^{\alpha+\beta}},
\fe
which gives us 
\ie 
\frac{-\epsilon \bar{w}^2 +|\epsilon|^2\bar{w} +\bar{\epsilon} |z|^2}{(|w|^2+|z|^2)^2(|w-\epsilon|^2+|z|^2)^2}=\frac{\Gamma(4)}{\Gamma(2)\Gamma(2)} \int_0^1 dx \frac{x(1-x)(-\epsilon \bar{w}^2 +|\epsilon|^2\bar{w} +\bar{\epsilon} |z|^2)}{(x(|w-\epsilon|^2+|z|^2)+(1-x)(|w|^2+|z|^2))^4}.
\fe
It can be shown that $x(|w-\epsilon|^2+|z|^2)+(1-x)(|w|^2+|z|^2)= |w-x\epsilon|^2 +|z|^2 + x(1-x)|\epsilon|^2$.
The integral \eqref{integd3} can then be rewritten as 
\label{integd4}
\ie 
 &\frac{1}{4\pi } t^a\otimes t^b f_{abc} \int_{t,w,\bar{w},z,\bar{z}} A^c(t,w,z) \wedge d w \wedge d \bar{w}  \wedge dz \wedge d \bar{z} \partial_z\left(  a \right)
\fe
where 
\ie
(a)= \frac{1}{2\pi^2} \frac{\Gamma(4)}{\Gamma(2)\Gamma(2)} \int_0^1 dx \frac{x(1-x)(-\epsilon \bar{w}^2 +|\epsilon|^2\bar{w} +\bar{\epsilon} |z|^2)}{(|w-x\epsilon|^2 +|z|^2 + x(1-x)|\epsilon|^2)^4}.
\fe

We would now like to show that $\partial_z(a)$ is proportional to $\partial_z \delta^4(w,z)$ in the limit where $\epsilon \rightarrow 0$. This requires us to evaluate the integral of $(a)$:
\ie \label{inttr}
C:=\int dw \wedge d\bar{w} \wedge dz \wedge d\bar{z}(a)
\fe
To this end, we make the change of variables $w\rightarrow w + x\epsilon $, whereby \eqref{inttr} becomes 
\ie \label{inttr2}
\int dw \wedge d\bar{w} \wedge dz \wedge d\bar{z} \frac{1}{2\pi^2} \frac{\Gamma(4)}{\Gamma(2)\Gamma(2)} \int_0^1 dx \frac{x(1-x)(-\epsilon (\bar{w}+x\bar{\epsilon})^2 +|\epsilon|^2(\bar{w}+x\bar{\epsilon}) +\bar{\epsilon} |z|^2)}{(|w|^2 +|z|^2 + x(1-x)|\epsilon|^2)^4}.
\fe 
Terms proportional to $\bar{w}^2$ and $\bar{w}$ in the numerator do not contribute to the integral, and we are left with 
\ie \label{inttr3}
\int dw \wedge d\bar{w} \wedge dz \wedge d\bar{z} \frac{1}{2\pi^2} \frac{\Gamma(4)}{\Gamma(2)\Gamma(2)} \int_0^1 dx \frac{x(1-x)( x(1-x)|\epsilon|^2\bar{\epsilon} +\bar{\epsilon} |z|^2)}{(|w|^2 +|z|^2 + x(1-x)|\epsilon|^2)^4}.
\fe 
Using polar coordinates $w=r_w e^{i\theta_w }$ and $z=r_z e^{i\theta_z }$, whereby the respective measures become $\int dw d \bar{w}= -i2 \int_0^{\infty } r_w dr_w \int_0^{2\pi }d\theta_w $ and $\int dz d \bar{z}= -i2 \int_0^{\infty } r_z dr_z \int_0^{2\pi } d\theta_z $, we can then perform the integrals straightforwardly. Performing the $r_w$ and angular integrals gives us 
\ie 
-8\int_0^1 dx \int_0^{\infty} r_z dr_z \left( \frac{((x)(1-x))^2 |\epsilon|^2\overline{\epsilon}}{(r_z^2+x(1-x)|\epsilon|^2)^3} + \frac{x(1-x)\overline{\epsilon}r_z^2}{(r_z^2+x(1-x)|\epsilon|^2)^3}\right).
\fe
The integral over $r_z$ can then be evaluated to give 
\ie 
-\frac{4}{\epsilon}\int_0^1 dx =-\frac{4}{\epsilon}.
\fe
An analysis analogous to that presented above shows that the term with $(b)$ does not contribute, as it involves performing an integral of the form $\int d^2z (\bar{z} f(|z|)) $ which gives zero.

Hence, we find that \eqref{integg} is 
\label{integd5}
\ie 
 &\frac{\hbar }{\pi \epsilon } t^a\otimes t^b f_{abc} \int dt \partial_z A^c(t,w,\bar{w},z,\bar{z}), 
\fe
where we have restored the appropriate factor of $\hbar$.

\section{Feynman Diagram Evaluation  for Fusion of General Line Operators}\label{genline}

In this appendix, we shall evaluate a Feynman amplitude   of the form
\ie 
\begin{aligned}
-\frac{1}{4\pi}f_{abc} \frac{t^a_{0, m}}{m!} \frac{t^b_{0, n}}{n!} \int_{V_2} d z_2 d w_2 A^c_{e x t} & \int_{V_0} \delta^{(2)}\left(z_0\right) \delta^{(2)}\left(w_0\right) \partial_{w_0}^m P_{02}  \int_{V_1} \delta^{(2)}\left(z_1\right) \delta^{(2)}\left(w_1-\tilde{w}\right) \partial_{w_1}^n  P_{12},
\end{aligned}
\fe 
where wedge products between differential forms are now implied, where we have used the notation 
\ie 
P_{ij}= P(t_i-t_j,w_i-w_j,z_i-z_j),
\fe 
and where $\delta^{(2)}(z)\delta^{(2)}(w)$ is a differential 4-form delta function proportional to $dw d\bar{w} dz d\bar{z}$. For concision, we shall suppress the factor of $-\frac{1}{4\pi m! n!}f_{abc} t^a_{0, m} t^b_{0, n} $, and reintroduce it at the end of the computation.  

Let us first consider the $V_0$ integral,
\ie 
\int_{V_0} \delta^{(2)}\left(z_0\right) \delta^{(2)}\left(w_0\right) \partial_{w_0}^m  P_{02}
\fe
which can be written explicitly as
\ie
-(-1)^{2m}\frac{5}{2}\cdot\frac{7}{2}\cdots\frac{5+2m-2}{2}\frac{3}{4}\frac{1}{\pi}\int_{V_0}\D^{(2)}(z_0)\D^{(2)}(w_0)\bar w_2^m \frac{(\bar z_2d\bar w_2-\bar w_2d\bar z_2)dt_0}{\sqrt{t^2_{02}+|w_{02}|^2+|z_{02}|^2}^{5+2m}}.
\fe
 Performing the shift of variables $t_0 \rightarrow t_0+ t_2$, and integrating over $t_0$ we obtain 
\ie\label{1}
-(-1)^{2m}\frac{\Gamma(2+m)}{\pi}\frac{\bar w_2^m\bar z_2(\bar z_2d\bar w_2-\bar w_2d\bar z_2)}{(|w_2|^2+|z_2|^2)^{m+3}}.
\fe
Now, let us consider the $V_1$ integral, which is of the form
\ie
\int_{V_1}\D^{(2)}(z_1)\D^{(2)}(w_1-\tilde w)\pa^n_{w_1}P_{12}.
\fe
Explicitly, after a shift $t_1\rightarrow t_1 + t_2$, it is given by
\ie
-(-1)^{n}\frac{5}{2}\cdot\frac{7}{2}\cdots\frac{5+2n-2}{2}\frac{3}{4}\frac{1}{\pi}\int^{\infty}_{-\infty}dt_1\frac{(\bar{\tilde w}-\bar w_2)^{n}(\bar z_2d\bar w_2+(\bar{\tilde w}-\bar w_2)d\bar z_2)}{\sqrt{t_1^2+|\tilde w-w_2|^2+|z_2|^2}^{5+2n}}.
\fe
Performing the integral over $t_1$, we find
\ie\label{2}
-(-1)^{n}\frac{\Gamma(2+n)}{2\pi}\frac{(\bar{\tilde w}-\bar w_2)^{n}(\bar z_2d\bar w_2+(\bar{\tilde w}-\bar w_2)d\bar z_2)}{(|\bar{\tilde w}-w_2|^2+|z_2|^2)^{n+2}}.
\fe
Combining both \eqref{1} and \eqref{2}, we can set up the $V_2$ integral: 
\ie
(-1)^{n} \frac{\Gamma(2+m)}{\pi}\frac{\Gamma(2+n)}{\pi}\int_{V_2}(dz_2dw_2)\frac{\bar w_2^m (\bar z_2d\bar w_2-\bar w_2d\bar z_2)}{(|w_2|^2+|z_2|^2)^{m+2}}\frac{(\bar{\tilde w}-\bar w_2)^{n}(\bar z_2d\bar w_2+(\bar{\tilde w}-\bar w_2)d\bar z_2)}{(|\bar{\tilde w}-w_2|^2+|z_2|^2)^{n+2}}A^c_{ext}.
\fe
We find that the nonvanishing contribution to the integral comes from expanding the external gauge field $A^c_{ext}$ to leading order in $z_2$:
\ie 
A^c_{ext}=\ldots+z_2\pa_{z_2}A^c_{ext},
\fe
whereby the integral can be expressed as 
\ie
(-1)^{n} \frac{\Gamma(2+m)}{\pi}\frac{\Gamma(2+n)}{\pi}\int_{V_2}\frac{\bar w_2^m(\bar{\tilde w}-\bar w_2)^{n}\bar z_2 \bar{\tilde w}(z_2 \pa_{z_2} A^c_{ext})}{(|w_2|^2+|z_2|^2)^{m+2}(|\tilde w-w_2|^2+|z_2|^2)^{n+2}}|dw_2|^2|dz_2|^2.
\fe
We now use the Feynman integral formula \eqref{feyn} to express the integral as 
\ie
(-1)^{n} \frac{\Gamma(4+m+n)}{(\pi)^2} \int^1_0 dx x^{m+1}(1-x)^{n+1}\int_{V_2}\frac{\bar w_2^m(\bar{\tilde w}-\bar w_2)^{n}\bar z_2 \bar{\tilde w}(z_2 \pa_{z_2} A^c_{ext})|dw_2|^2|dz_2|^2}{((1-x)(|w_2|^2+|z_2|^2)+x(|\tilde w-w_2|^2+|z_2|^2))^{m+n+4}}.
\fe
The denominator of this formula can be reexpressed as 
$\left(\left|w_2-x \tilde{w}\right|^2+\left|z_2\right|^2+x(1-x)|\tilde{w}|^2\right)^{m+n+4}$. Performing the shift $w_2 \rightarrow w_2 +x \tilde{w}$, we obtain
\ie
\tilde{c}_{mn}\int^1_0 dx x^{m+1}(1-x)^{n+1}\int_{V_2}\pa_{z_2} A^c_{ext}\frac{(\bar w_2+x\bar{\tilde w})^m((1-x)\bar{\tilde w}-\bar w_2)^{n}|z_2|^2\bar{\tilde w}}{(|w_2|^2+|z_2|^2+x(1-x)|\tilde w|^2)^{m+n+4}}|dw_2|^2|dz_2|^2,
\fe
where we have defined
\ie 
\tilde{c}_{mn} = (-1)^{n} \frac{\Gamma(4+m+n)}{(\pi)^2}. 
\fe 
Due to the integral over the angular component of $w_2$, the terms in the numerator proportional to powers of $\bar{w}_2$ do not contribute, and we find that it takes the form
\ie
\tilde{c}_{mn}\bar{\tilde w}^{m+n+1}\int^1_0 dx x^{m+1}(1-x)^{n+1}\int_{V_2}\pa_{z_2} A^c_{ext}\frac{|z_2|^2}{(|w_2|^2+|z_2|^2+x(1-x)|\tilde w|^2)^{m+n+4}}|dw_2|^2|dz_2|^2 .
\fe
Now, explicitly using the polar coordinates $z=r_ze^{i\theta_z}$ and $w=r_we^{i\theta_w}$, and integrating over $\theta_z$ and $\theta_w$, we find 
\ie
&-\tilde{c}_{mn} \bar{\tilde w}^{m+n+1}\int^1_0dx x^{m+1}(1-x)^{n+1}\int_{\bR_t}\pa_{z_2}A^c_{ext}\int^\infty_0\int^\infty_0\frac{16\pi^2r^3_zr_w}{(r_z^2+r_w^2+x(1-x)|\tilde w|^2)^{m+n+4}}dr_zdr_w\\
=~&-\tilde{c}_{mn} \frac{4\pi^2}{\prod_{l=1}^3(l+m+n)}{\tilde w}^{-m-n-1}\int_{\bR_t}\pa_{z_2}A^c_{ext}\int^1_0 dx\\
=~&-\tilde{c}_{mn} \frac{4\pi^2}{\prod_{i=1}^3(l+m+n)}{\tilde w}^{-m-n-1}\int_{\bR_t}\pa_{z_2} A^c_{ext}.
\fe

\section{Feynman Diagram Evaluation  for Fusion of Surface Operators }\label{surffuse}

In this appendix, we would like to evaluate the Feynman amplitude \eqref{surffuseampl}.
To evaluate the integrals over $w_1$ and $w_2$, we ought to first shift the integration variables as $w_1 \rightarrow w_1 +w$ and $w_2 \rightarrow w_2 +w$, leading to 
\ie \label{jjint}
&-\frac{1}{(-i2)^2}\int_{w_1}  \int_{w_2} \int_{t,w,\bar{w},z,\bar{z}}\frac{\hbar}{4\pi}  J^a(w_1 + w)\otimes \tilde{J}^b(w_2+ w) f_{abc}\\&\times   P(t,w_1,z)\wedge dw\wedge dz \wedge A_{ext}^c(t,w,z)\wedge P(t-\epsilon,w_2,z)
\fe 
Taylor expanding the currents as $J^a(w_1+w)=\sum_{m\in \{0\} \cup \mathbb{N}}J_m^a(w_1+w)^m$ and $\tilde{J}^a(w_2+w)=\sum_{m\in \{0\} \cup \mathbb{N}}\tilde{J}_n^a(w_2+w)^n$, and performing a binomial expansion of each power of $(w_1+ w)$ and $(w_2+w)$, we find that each term contributes $w^m$ and $w^n$ respectively due to the integrals over $w_1$ and $w_2$. 

The integral \eqref{jjint} can then be written as 
\ie
-\frac{1}{(-i2)^2}\int_{w_1}  \int_{w_2} \int_{t,w,\bar{w},z,\bar{z}}\frac{\hbar}{4\pi}  J^a(w)\otimes \tilde{J}^b( w) f_{abc}  P(t,w_1,z)\wedge dw\wedge dz \wedge A_{ext}^c(t,w,z)\wedge P(t-\epsilon,w_2,z).
\fe 
Using the fact that 
\ie 
\int r_w dr_w d\theta_w \frac{1}{(r_w^2 +|z|^2 +t^2)^\frac{5}{2}} = 2\pi \frac{1}{3} \frac{1}{(|z|^2 +t^2)^{\frac{3}{2}}},
\fe
we can compute the integrals over $w_1$ and $w_2$, to give 
\ie \label{penul}
-\frac{\hbar}{4\pi }\int_{t,w,\bar{w},z,\bar{z}} J^a(w)\otimes \tilde{J}^b(w) f_{abc} dw \wedge A_{ext}^c(t,w,z) \wedge dz \wedge P'(t,z,\bar{z}) \wedge P'(t-\epsilon,z,\bar{z}),
\fe 
where 
\ie \label{pprime2}
P^{\prime}(t, z, \bar{z}):= -\frac{1}{4}\left(-\mathrm{d} \bar{z} \frac{\partial}{\partial t}+4 \mathrm{~d} t \frac{\partial}{\partial z}\right) \frac{1}{\left(t^2+z \bar{z}\right)^{\frac{1}{2}}}.
\fe 
The quantity \eqref{pprime2} is familiar from derivation of the fusion of Wilson lines in 4d Chern-Simons theory in \cite{Costello:2017dso}, where it was shown that 
\ie 
\lim _{\epsilon \rightarrow 0} \mathrm{~d} z \wedge P^{\prime}(t, z, \bar{z}) \wedge P^{\prime}(t-\epsilon, z, \bar{z})=\frac{\pi}{\mathrm{i}} \frac{\partial}{\partial z} \delta^3(t, z, \bar{z}).
\fe 

\section{Feynman Diagram Evaluation for Line-Surface Coproduct}\label{linesurf}

We would like to evaluate the Feynman amplitude
\ie 
&-\frac{\hbar}{4\pi} \left(\frac{1}{-i2}\right)\sum_n J^a_{-n-1}\frac{t^{b}_{0,m}}{m!}f_{abc}\int_{V_2}A_{ext}^c (t_2,w_2,z_2)\wedge dw_2 \wedge dz_2 \wedge \int_{V_0} \delta(t_0) \delta^{(2)}(z_0) w_0^n P_{02} dw_0 \\ &\wedge \int_{V_1} \delta (t_1+\epsilon) \delta^{(2)}(z_1) \delta^{(2)}(w_1)\partial^m_{w_1} P_{12},
\fe
where the location of the vertex $V_1$ has been specified to be $t_1=-\epsilon$.
Here, we have included a factor of $(-1/i2)$, since the measure of our surface operator is defined as $\frac{dw d\bar{w}}{-i2}$.
Let us first consider the $V_0$ integral, which takes the form
\ie 
\frac{3}{4}\frac{1}{2\pi}\int_{\mathbb{C}_{w_0}} \frac{w_0^n (2\bar{z}_2 dt_2 - t_2 d\bar{z}_2)}{(t_2^2 +|z_2|^2 +|w_{02}|^2)^{\frac{5}{2}}} d^2w_0
\fe 
Making the shift of variable $w_0 \rightarrow w_0 + w_2 $, we have
\ie 
&\frac{3}{4}\frac{1}{2\pi} \int_{\mathbb{C}_{w_0}} \frac{(w_0+w_2)^n (2\bar{z}_2 dt_2 -t_2 d\bar{z}_2)}{(t_2^2 +|z_2|^2 +|w_{0}|^2)^{\frac{5}{2}}} d^2w_0\\=&\frac{3}{4}\frac{1}{2\pi}\int_{\mathbb{C}_{w_0}} \frac{(w_2)^n (2\bar{z}_2 dt_2 - t_2 d\bar{z}_2)}{(t_2^2 +|z_2|^2 +|w_{0}|^2)^{\frac{5}{2}}} d^2w_0,
\fe
since the integral over the angular coordinate of $w_0$ of positive powers of $w_0$ vanishes. 
Using the polar coordinates $w_0=r_{w_0} e^{i\theta_{w_0}}$, we find 
\ie \label{exa}
- \frac{1}{4} \frac{w_2^n (2\bar{z}_2dt_2 -t_2 d\bar{z}_2)}{(|z_2|^2 +t_2^2)^{\frac{3}{2}}}.
\fe

The $V_1$ integral
\ie 
\int_{V_1}\delta (t_1+\epsilon )\delta^{(2)}(z_1)\delta^{(2)}(w_1)\partial_{w_1}^m P_{12}
\fe 
is straightforward due to the presence of the three delta functions. It can be evaluated to be
\ie \label{exb}
\frac{3}{4}\frac{1}{2\pi}\left((-1)^{2m} \frac{5}{2}\frac{7}{2} \ldots \frac{5+2m-2}{2}\right)\frac{(\bar{w}_2^m (-2\bar{z}_2 d\bar{w}_2 dt_2 + 2\bar{w}_2 d\bar{z}_2 dt_2 +(\epsilon + t_2)d\bar{z}_2 d\bar{w}_2))}{(((\epsilon+t_2)^2)+|z_2|^2+|w_2|^2)^{\frac{5+2m}{2}}}.
\fe

Setting $m=n$ and combining the expressions \eqref{exa} and \eqref{exb} while suppressing numerical coefficients that will be restored at the end, we can set up the $V_2$ integral
\ie 
\int_{V_2} dw_2 dz_2 \frac{ |w_2|^{2n} (2\bar{z}_2dt_2 -t_2 d\bar{z}_2)}{(|z_2|^2 +t_2^2)^{\frac{3}{2}}}\frac{((2\bar{z}_2 d\bar{w}_2 dt_2 -2\bar{w}_2 d\bar{z}_2 dt_2 +(\epsilon + t_2)d\bar{z}_2 d\bar{w}_2))}{((\epsilon+t_2)^2+|z_2|^2+|w_2|^2)^{\frac{5+2n}{2}}}A^c_{ext}.
\fe
Expanding the external gauge field $A^c_{ext}$, we notice that the nonvanishing part of the integral arises from one of its modes :
\ie 
A^c_{ext}(z_2) = \ldots + z_2\partial_{z_2} A^c_{ext}.
\fe
The $V_2$ integral can then be written as 
\ie 
\int_{V_2} \frac{\left|w_2\right|^{2 n}\left|z_2\right|^2\left(\epsilon\right)}{{({t_2^2+\left|z_2\right|^2})}^{\frac{3}{2}} ({\left(\epsilon+t_2\right)^2+\left|w_2\right|^2+\left|z_2\right|^2})^{\frac{5}{2}+n}} d t_2\left|d w_2\right|^2\left|d z_2\right|^2.
\fe 

Using \eqref{feyn}, and setting $\epsilon=1$, we find (suppressing coefficients involving gamma functions)
\ie 
\int_0^1 dx\int_{V_2} \frac{\left|w_2\right|^{2 n}\left|z_2\right|^2 x^{\frac{3}{2}+n}(1-x)^{\frac{1}{2}}}{(x((1+t_2)^2 +|z_2|^2 +|w_2|^2) +(1-x)(|z_2|^2 +t_2^2))^{4+n}} d t_2\left|d w_2\right|^2\left|d z_2\right|^2
\fe 
This can be shown to be equivalent to 
\ie 
\int_0^1 dx\int_{V_2} \frac{\left|w_2\right|^{2 n}\left|z_2\right|^2 x^{\frac{3}{2}+n}(1-x)^{\frac{1}{2}}}{(|z_2|^2+x|w_2|^2 +(t_2+x)^2 +x(1-x))^{4+n}} d t_2\left|d w_2\right|^2\left|d z_2\right|^2.
\fe 
Making the shift $t_2\rightarrow t_2 -x$, we find 
\ie 
\int_0^1 dx\int_{V_2} \frac{\left|w_2\right|^{2 n}\left|z_2\right|^2 x^{\frac{3}{2}+n}(1-x)^{\frac{1}{2}}}{(|z_2|^2+x|w_2|^2 +(t_2)^2 +x(1-x))^{4+n}} d t_2\left|d w_2\right|^2\left|d z_2\right|^2.
\fe

Using polar coordinates, and computing the integrals over angular variables, the remaining integral is
\ie 
(-i2)^2 4\pi^2\int_0^1 dx\int \frac{r_w^{2n+1}r_z^3  (x^{\frac{3}{2}+n})(1-x)^{\frac{1}{2}}}{(r_z^2 + x r_w^2 + t_2^2 +x(1-x))^{4+n}}dt_2 dr_w dr_z.
\fe 
Performing the rescaling $r_w \rightarrow r_w/\sqrt{x}$, the integral becomes 
\ie 
&(-i2)^2  4\pi^2\int_0^1 dx\int  \frac{r_w^{2n+1}r_z^3  (x^{\frac{1}{2}})(1-x)^{\frac{1}{2}}}{(r_z^2 +  r_w^2 + t_2^2 +x(1-x))^{4+n}}dt_2 dr_w dr_z.
\fe 
We can then evaluate the remaining integrals which give 
\ie 
(-i2)^2 4\pi^2  \frac{\pi}{(4+4n) (6+5n +n^2)}.
\fe 

Let us now address the overall numerical coefficient that has been suppressed up to this point. In particular, we are concerned with the dependence on $n$ in this coefficient. The coefficient is 
\ie
\left(-\frac{\hbar}{8\pi }\frac{1}{n!}  \frac{3}{8}\right)\frac{5}{2}\frac{7}{2} \ldots \frac{5+2n-2}{2}\frac{\Gamma(4+n)}{\Gamma(\frac{3}{2})\Gamma(\frac{5}{2}+n)}\frac{1}{(n+3)(n+2)(n+1)}.
\fe 
Here $\frac{5}{2}\frac{7}{2} \ldots \frac{5+2n-2}{2}$ arises from \eqref{exb}, the ratio of gamma functions arises from using the Feynman integral identity \eqref{feyn}, $1/(n+3)(n+2)$ arises from the integral over $r_{z_2}$, and $1/(n+1)$ arises from the integral over $r_{w_2}$.
Since 
\ie 
\frac{5}{2}\frac{7}{2} \ldots \frac{5+2n-2}{2}\frac{\Gamma(4+n)}{\Gamma(\frac{5}{2}+n)}=\frac{4}{3}\frac{\Gamma(4+n)}{\sqrt{\pi}}
\fe
and 
\ie 
\frac{\Gamma(4+n)}{(n+3)(n+2)(n+1)}=n!,
\fe
which cancels the factor of $1/n!$ coming from the coefficient of the coupling of the gauge field to $t^b_{0,n}$, we find that the $n$-dependence cancels out in the final numerical coefficient, as expected. 
\bibliographystyle{ytphys}
\bibliography{5dCS}

\end{document}